\documentclass[]{aa}  

\usepackage{graphicx,natbib,color}
\usepackage{txfonts}
\usepackage[]{hyperref}
\begin{document} 

% adaptive
   \title{Inference of magnetic fields in the very quiet Sun}

   \subtitle{}

   \author{M. J. Mart\'{\i}nez Gonz\'alez
          \inst{1,2}
          \and
          A. Pastor Yabar\inst{1,2}
          \and A. Lagg\inst{3}
          \and A. Asensio Ramos\inst{1,2}
          \and M. Collados\inst{1,2}
          \and S. K. Solanki\inst{3,4}
          \and \hbox{H. Balthasar\inst{5}}
          \and T. Berkefeld\inst{6}
          \and C. Denker\inst{5}
          \and H. P. Doerr\inst{3}
          \and A. Feller\inst{3}
          \and M. Franz\inst{6}
          \and S. J. Gonz\'alez Manrique\inst{5,7}
          \and A. Hofmann\inst{5}
          \and F. Kneer\inst{8}
          \and \hbox{C. Kuckein\inst{5}}
          \and R. Louis\inst{5}
          \and O. von der L\"uhe\inst{6}
          \and H. Nicklas\inst{3}
          \and D. Orozco\inst{1,2}
          \and R. Rezaei\inst{1,2,6}
          \and R. Schlichenmaier\inst{6}
          \and \hbox{D. Schmidt\inst{6}}
          \and W. Schmidt\inst{6}
          \and M. Sigwarth\inst{6}
          \and M. Sobotka\inst{9}
          \and D. Soltau\inst{6}
          \and J. Staude\inst{5}
          \and K. G. Strassmeier\inst{5}
          \and M. Verma\inst{5}
          \and \hbox{T. Waldman\inst{6}}
          \and R. Volkmer\inst{6}
          }

    \institute{Instituto de Astrof\'\i sica de Canarias, V\'\i a L\'actea s/n, E-38205 La Laguna, Tenerife, Spain
         \and
              Dept. Astrof\' isica, Universidad de La Laguna, E-38205, La Laguna, Tenerife, Spain
         %\and Max Planck Institute for Solar System Research, Justus-von-Liebig-Weg 3, 37077, G\"ottingen, Germany
         \and Max-Planck-Institut f\"ur Sonnensystemforschung, Justus-von-Liebig Weg 3, 37077 G\"ottingen, Germany
         \and School of Space Research, Kyung Hee University,Yongin, Gyeonggi-Do, 446-701, Republic of Korea
         \and Leibniz-Institut f\"ur Astrophysik Potsdam (AIP), An der Sternwarte 16, 14482 Potsdam, Germany
         \and Kiepenheuer Institut f\"ur Sonnenphysik, Sch\"oneckstr. 6, 79104 Freiburg, Germany
         \and Universit\"at Potsdam, Institut  f\"ur Physik und Astronomie, Karl-Liebknecht-Strasse 24/25, 14476 Potsdam-Golm, Germany
         \and Institut f\"ur Astrophysik, Friedrich Hund Platz 1, 37077 G\"ottingen, Germany
         \and Astronomical Institute, Academy of Sciences of the Czech Republic, Fri\v{c}ova 298, 25165 Ond\v{r}ejov, Czech Republic}

   \date{}

 \abstract{Over the past 20 years, the quietest areas of the solar surface have revealed a weak but extremely dynamic magnetism occurring at small scales ($< 500$ km), which may provide an important contribution to the dynamics and energetics of the outer layers of the atmosphere. Understanding this magnetism requires the inference of physical quantities from high-sensitivity
spectro-polarimetric
data with high spatio-temporal resolution. }
 {We present high-precision spectro-polarimetric data with high spatial resolution (0.4$''$) of the very quiet Sun at 1.56$\mu$m obtained with the GREGOR telescope to shed some light on this complex magnetism. }
 {We used inversion techniques in two main approaches. First, we assumed that the observed profiles can be 
 reproduced with a constant magnetic field atmosphere embedded in a field-free medium. Second, we assumed that the resolution element has a substructure with either two constant magnetic atmospheres or a single magnetic atmosphere with gradients of the physical quantities along the optical depth, both coexisting with a global stray-light component.}
 {Half of our observed quiet-Sun region is better explained by magnetic substructure within the resolution element. However, we cannot distinguish whether this substructure comes from gradients of the physical parameters along the line of sight or from horizontal gradients (across the surface). In these pixels, a
model with two magnetic components is preferred, and we find two distinct magnetic field populations. The population with the larger filling factor has very weak ($\sim$150 G) horizontal fields similar to those obtained in previous works. We demonstrate that the field vector of this population is not constrained by the observations, given the spatial resolution and polarimetric accuracy of our data. The topology of the other component with the smaller filling factor is constrained by the observations for field strengths above 250 G: we infer hG fields with inclinations and azimuth values compatible with an isotropic distribution. The filling factors are typically below 30\%. We also find that the flux of the two polarities is not balanced. 
 From the other half of the observed quiet-Sun area $\sim$50\% are two-lobed Stokes $V$ profiles, meaning that 23\% of the field
of view  can be adequately explained with a single constant magnetic field embedded in a non-magnetic atmosphere. The magnetic field vector and filling factor are reliable inferred in only 50\% based on the regular profiles. Therefore, 12\% of the field of
view harbour hG fields with filling factors typically below 30\%. At our present spatial resolution, 70\% of the pixels apparently
are non-magnetised.}
{}
%5 {} token are mandatory
 
  % \abstract{}


   \maketitle
%
%________________________________________________________________

\section{Introduction}

At any given time, even at the maximum of the Sun's activity cycle, 
most of the solar surface is covered by areas of low average magnetic 
fluxes. These areas are called the quiet Sun. 
Within the quiet Sun, we refer to the network as the reticular pattern at the 
border of supergranular cells, which has magnetic fluxes of $\sim 10^{18} - 10^{19}$ Mx \citep[e.g.][]{stenflo_73, yo_12_network}. 
The interiors of these cells are permeated by a weaker magnetism whose fluxes are lower by 
one or two orders of magnitude. We call these regions the very quiet Sun. 
Recently, evidence of even more quiet areas within the very quiet Sun has been reported, 
the so-called dead calm areas, where magnetic fluxes are the weakest detected \citep[$\sim 10^{15} Mx$;][]{yo_12}.

Very sensitive spectro-polarimeters are needed to detect the very faint
polarisation signals  (more than two orders of magnitude weaker than those
from active regions) of the very quiet areas of the Sun.  In addition, high 
spatio-temporal resolution observations are needed to properly study the small-scale
dynamic magnetism of these regions.  The spectro-polarimeter of the Hinode
satellite \citep{hinode_sp}, the Imaging Magnetograph eXperiment \citep[IMaX]{imax}
instrument onboard the Sunrise mission \citep{sunrise}, and the Tenerife
Infrared Polarimeter \citep{tipII} using the adaptive optics system \citep{AO} of the German
Vacuum Tower Telescope (Observatorio del Teide) have provided such data during
the past decade  \citep[e.g.][and references
therein]{khomenko_03,arturo06,david_07,lites_08,yo_08,sanja_10,carlosq_13,iker_15}.
This explains why the magnetism of the quiet Sun has received increasing
attention in the past few years. This is  also partly motivated by the exceptionally
extended minimum following solar cycle 23 (2007-2011).

\begin{figure*}[!ht]
\centering
\includegraphics[width=\textwidth, viewport= 30 42 450 126]{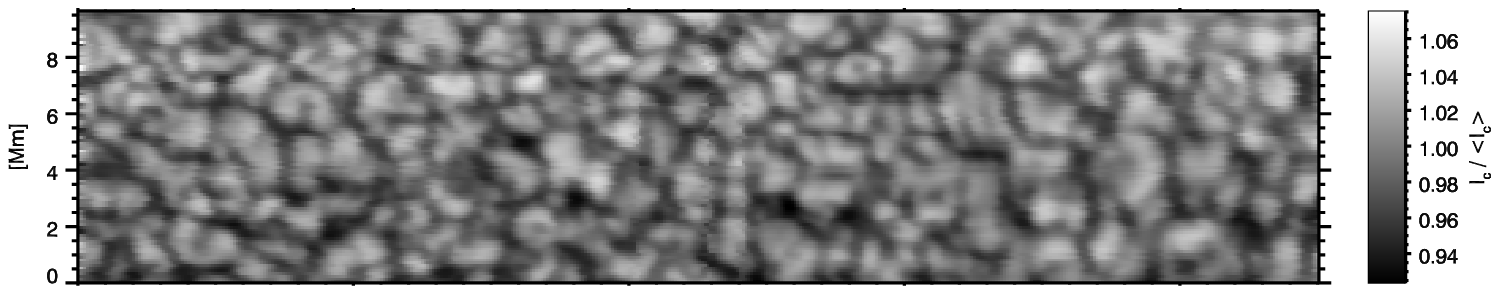}
\includegraphics[width=\textwidth, viewport= 30 42 450 126]{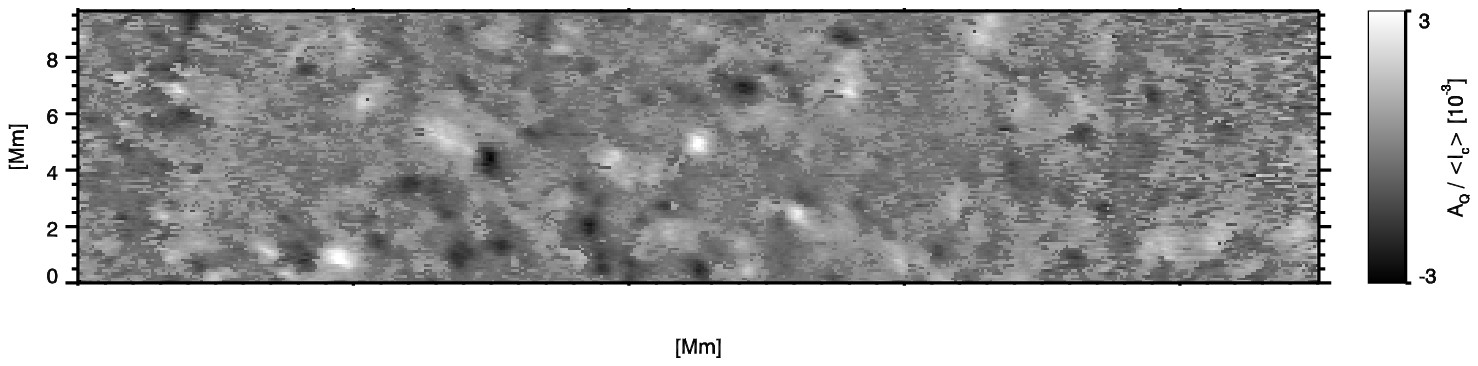}
\includegraphics[width=\textwidth, viewport= 30 42 450 126]{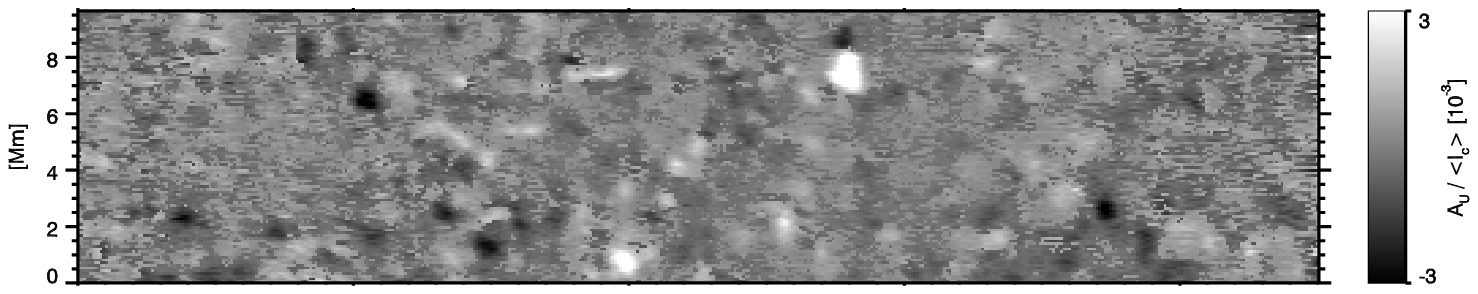}
\includegraphics[width=\textwidth, viewport= 30 8 450 126]{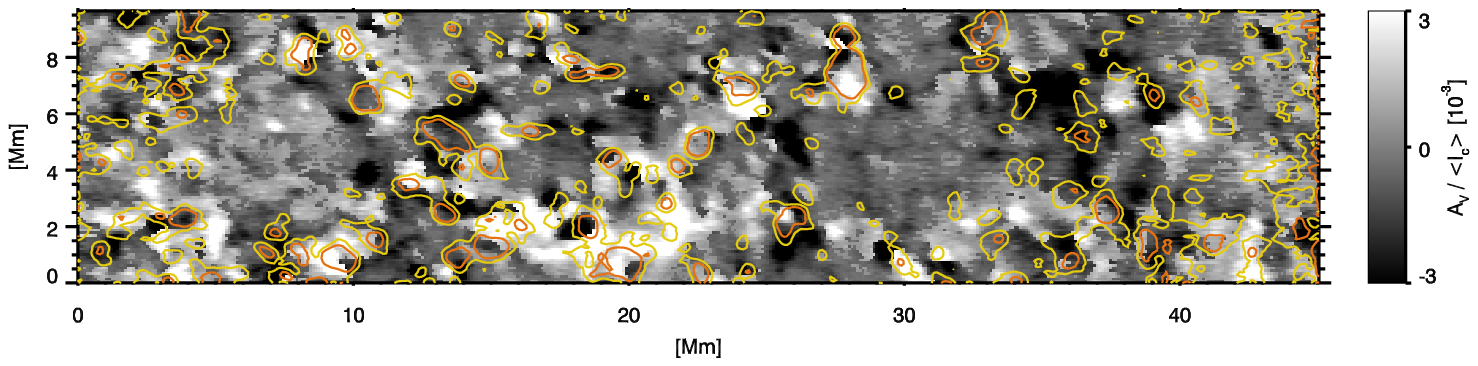}
\caption{Intensity and polarisation maps of the observed quiet region at disc centre. The yellow and 
orange contours display the linear polarisation $\sqrt{Q^2+U^2}$ with amplitude levels of 1$\times 10^{-3}$ 
$\langle$I$_\mathrm{c}\rangle$ and 1.5$\times 10^{-3}$ $\langle$I$_\mathrm{c}\rangle$, respectively. Although the 
defined linear polarisation is a biased quantity, these contours are reliable since their values are more than one order of magnitude 
above the noise level. The amplitude 
of the circular polarisation has been computed as the highest value of the amplitudes of the two lobes, and the sign is given by the blue lobe. 
The amplitude of a lobe is calculated as the average value in two pixels around the maximals. The amplitudes of Stokes $Q$ or $U$ are calculated as the average in two pixels around the position of the maximum, taking the absolute value of the parameters.}
\label{mapas}
\end{figure*}

\begin{figure*}
\centering
\includegraphics[viewport=30 42 450 126, width=\textwidth]{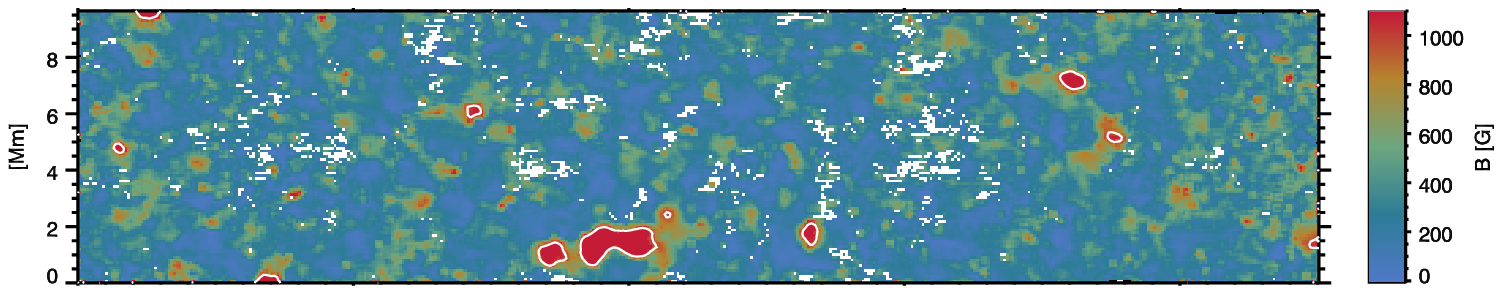}
\includegraphics[viewport=30 42 450 126, width=\textwidth]{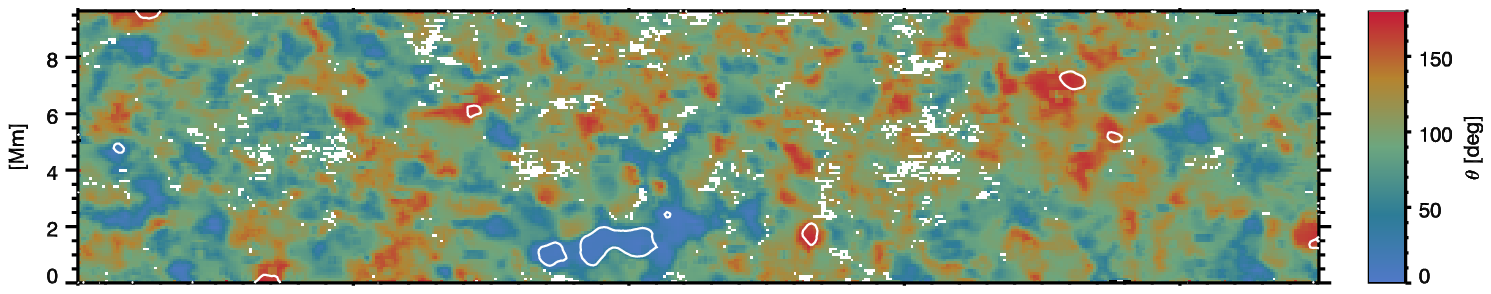}
\includegraphics[viewport=30 42 450 126, width=\textwidth]{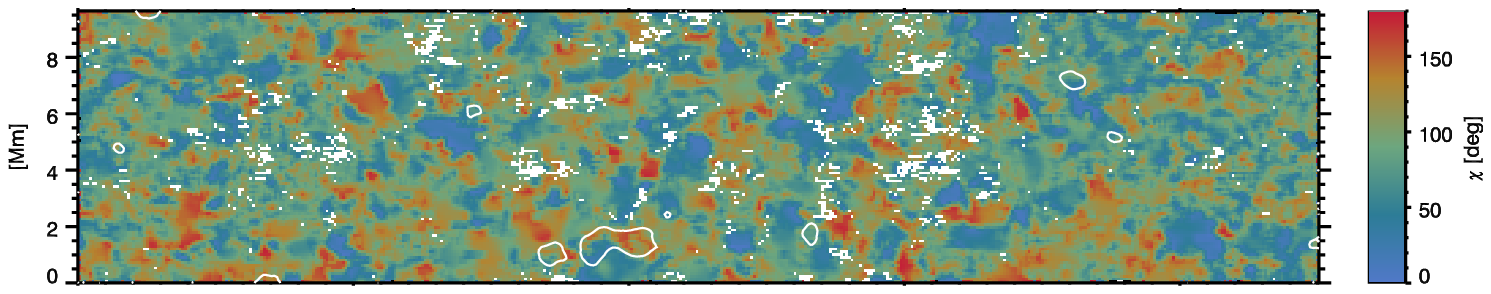}
\includegraphics[viewport=30 8 450 126, width=\textwidth]{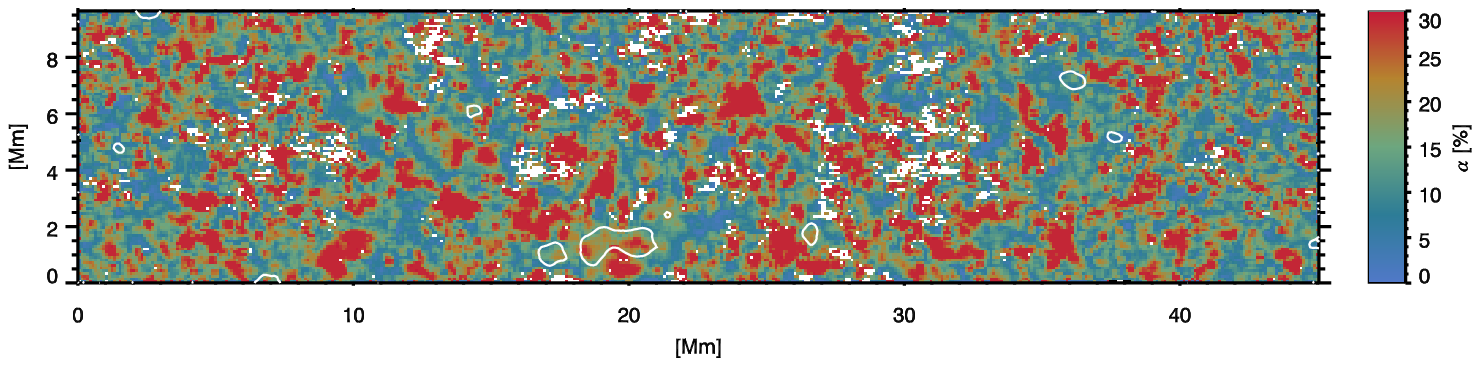}
\caption{Maps of the inferred magnetic field vector (strength, inclination, and azimuth) and the filling factor 
of the magnetic component. The white contours denote the areas with kG fields. White pixels contain signal below $4\sigma_n$ and hence are not 
inverted.}
\label{mapas_inv}
\end{figure*}

\begin{figure}
\centering
\includegraphics[width=\columnwidth,viewport=43 9 493 336]{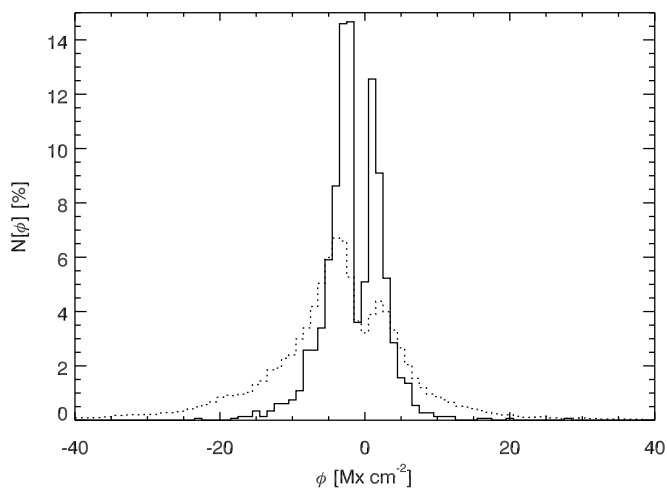}
\caption{Histograms of the magnetic flux density inferred using a model with one magnetic atmosphere 
embedded in a field-free volume. The histograms represent the inversions at those pixels where Stokes $Q$ or $U$ or $V$ are above 4$\sigma_n$, model 1C is preferred by the observations, and the Stokes $V$ is a two-lobed profile. The solid line represents the values of the magnetic flux density for those pixels where the magnetic field strength, inclination, and filling factor are not reliably retrieved. The dotted line displays the histogram for those pixels where the magnetic field vector, inclination, and azimuth are well constrained by the observations.}
\label{hist_flujo}
\end{figure}

\begin{figure*}
\centering
\includegraphics[width=0.4\textwidth,viewport=43 9 493 336]{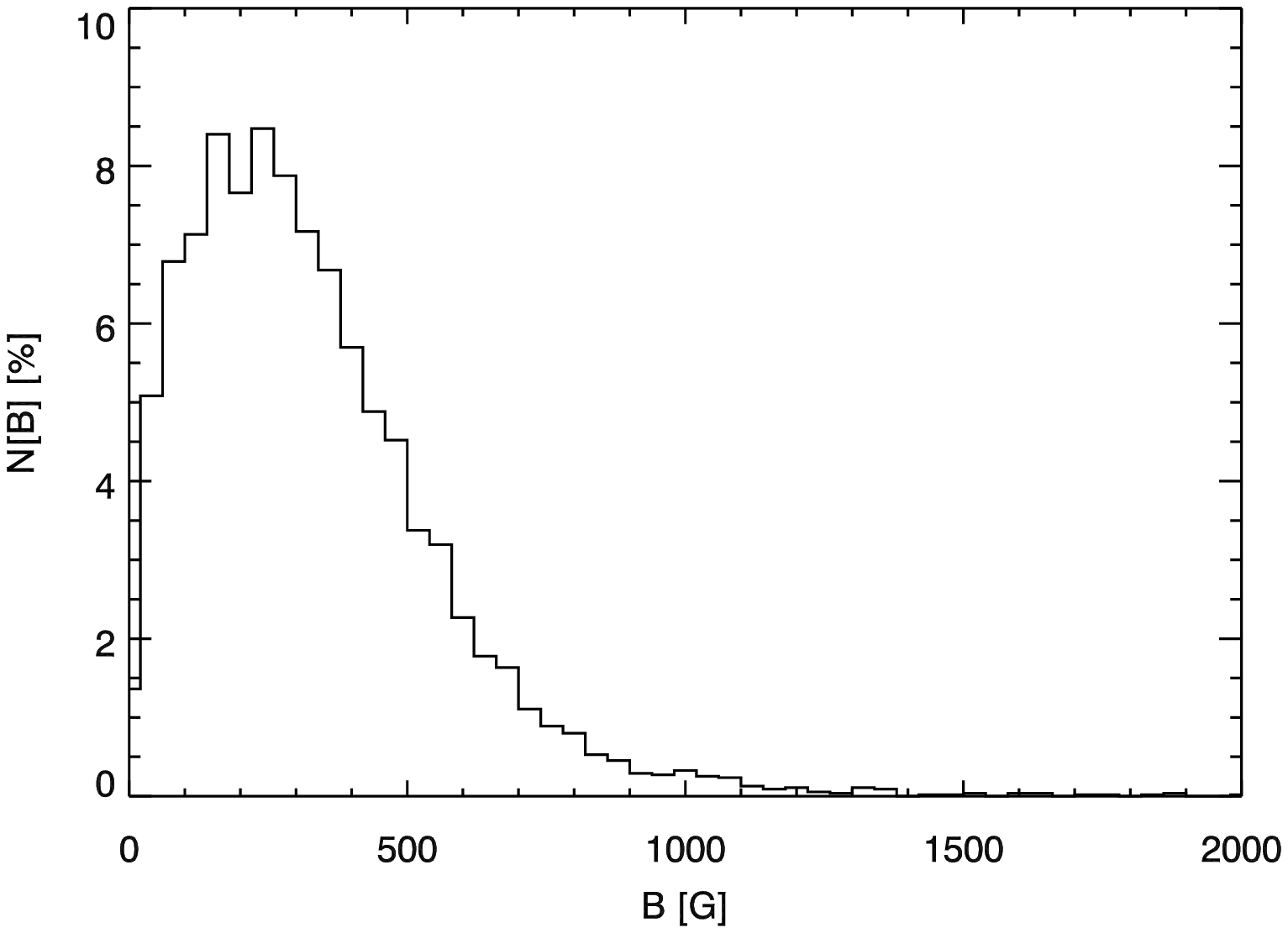}
\includegraphics[width=0.4\textwidth,viewport=43 9 493 336]{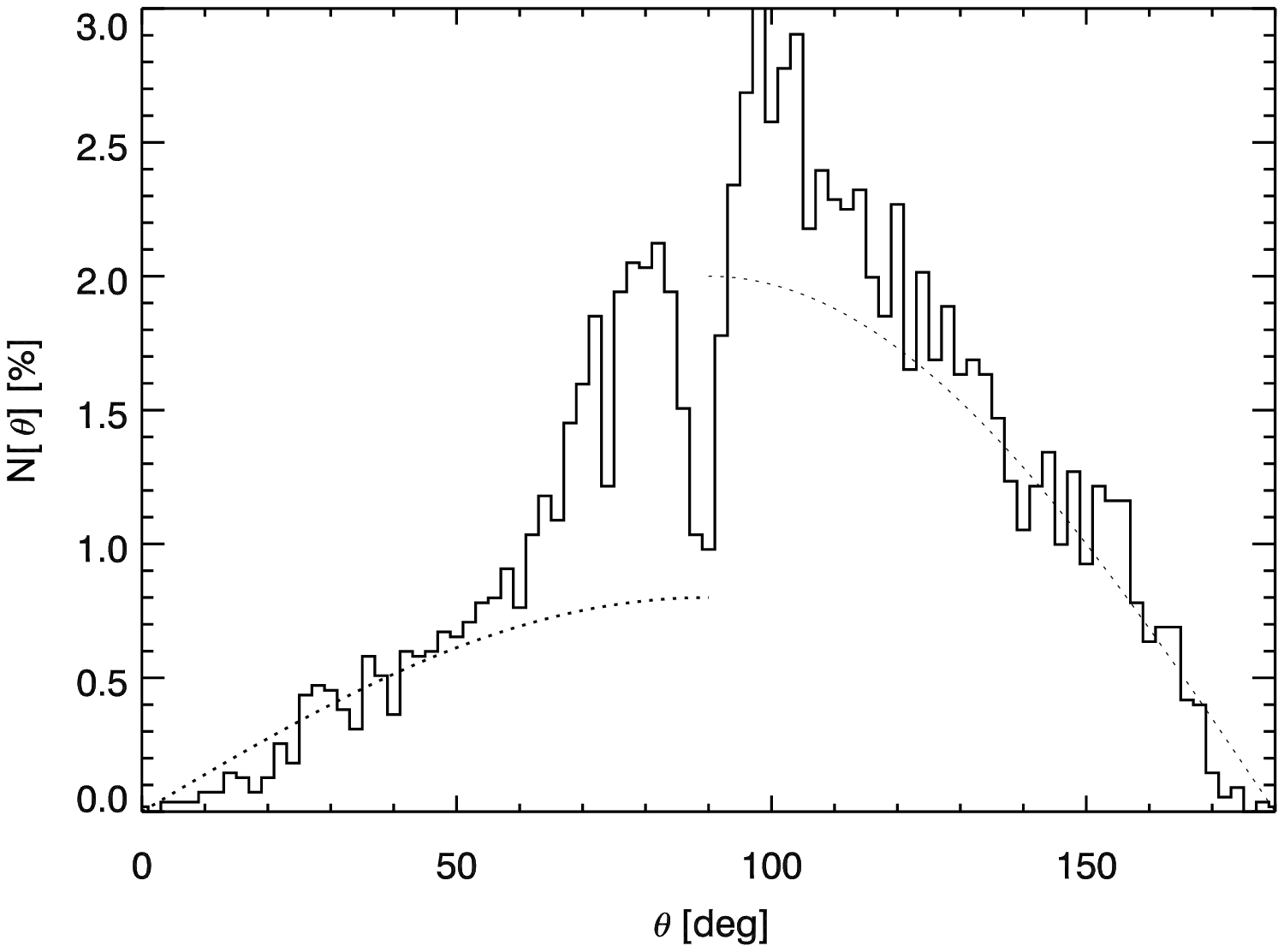}\\
\includegraphics[width=0.4\textwidth,viewport=43 9 493 336]{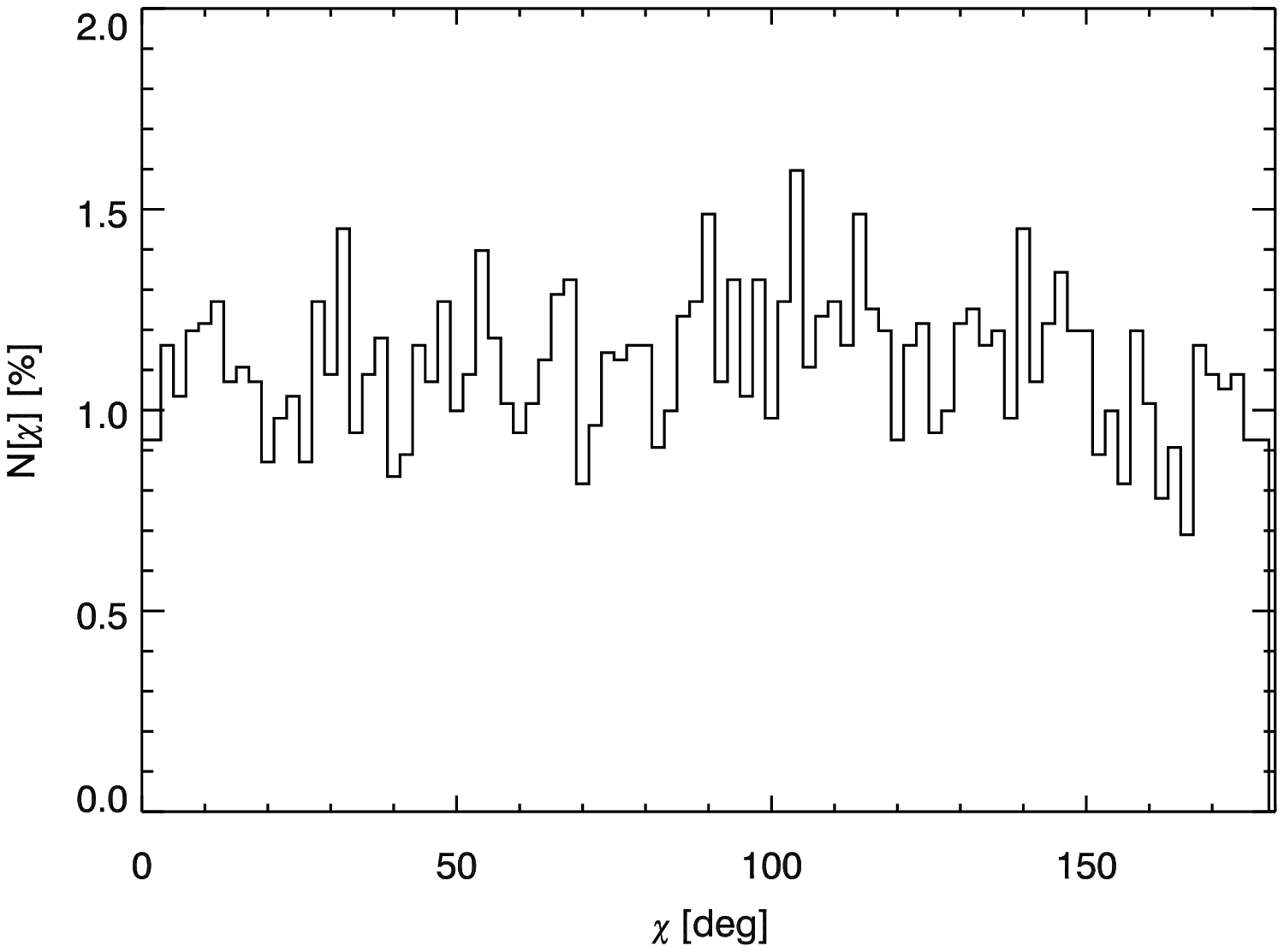}
\includegraphics[width=0.4\textwidth,viewport=43 9 493 336]{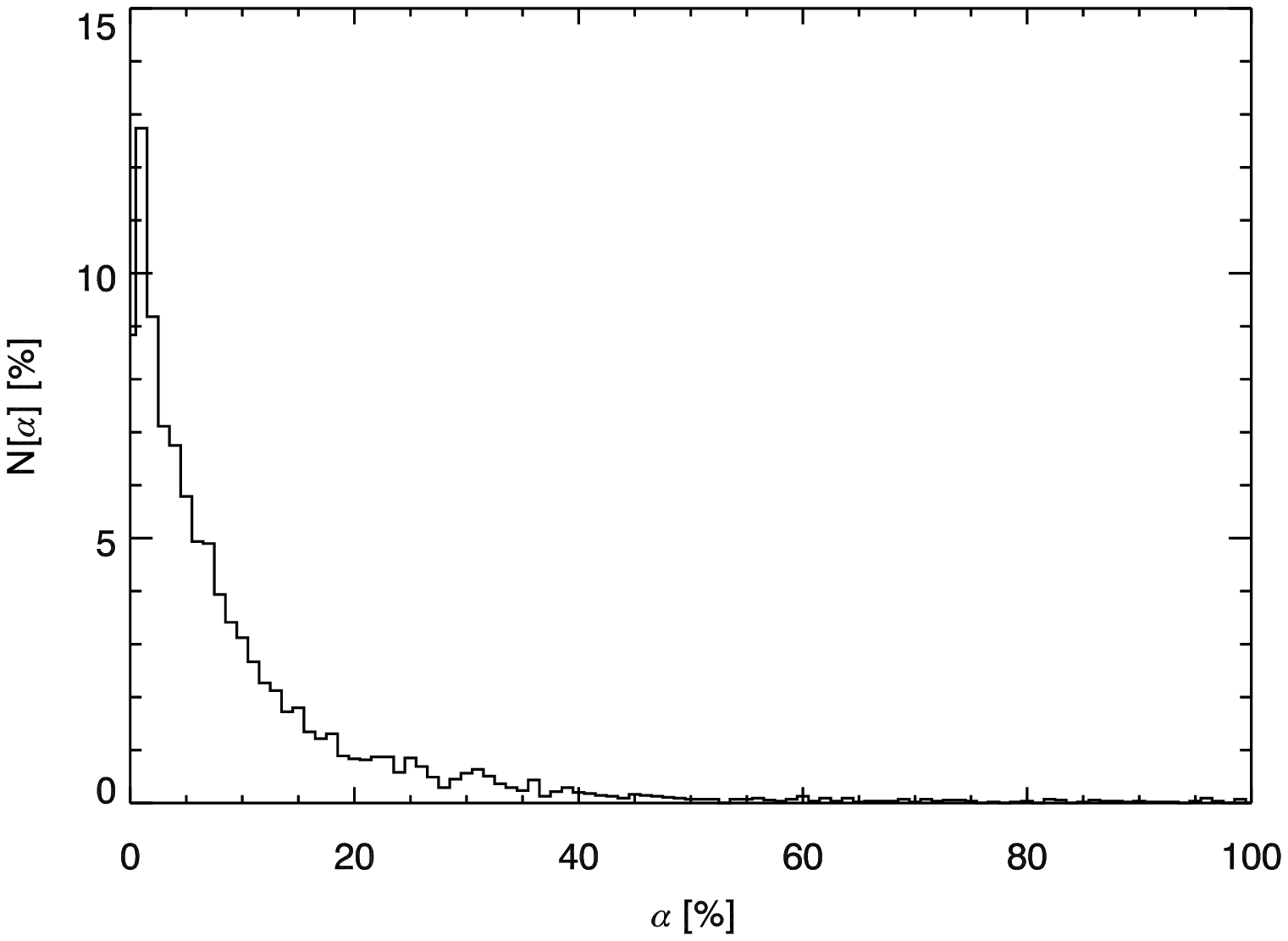}
\caption{Histograms of the inferred field vector (strength, inclination, and azimuth) and the filling factor 
of the magnetic component using a model with one magnetic atmosphere 
embedded in a field-free volume. The histograms represent the inversions at those pixels where Stokes $Q$ or $U$ or $V$ 
are above 4$\sigma_n$, the Stokes $V$ is a two-lobed profile, and the field vector and filling factor are well constrained by the observed profiles.}
\label{hist_inv}
\end{figure*}

When analysing the statistical properties of polarimetric signals, most of the works in the literature agree on the 
picture of the very quiet Sun magnetism: the signals are the same everywhere in the solar disc. In other words, the quiet Sun 
looks the same regardless of the observer’s line of sight \citep{yo_08_mus,lites_08,david_12}. Moreover, their magnetic properties either 
do vary with the activity cycle, or this variation is very weak \citep[][]{faurobert_01,buehler_13,lites_14,faurobert_15}. Finally, the degree of spatial coherence of the 
signals increases with the polarimetric signal. By this we mean that the weakest detected fluxes of the quiet Sun on scales below 1000 km show a very 
dynamic, intermittent, stochastic, magnetic activity unlike the "deterministic" characterisation of the magnetic structure of sunspots 
and other active structures formed by the so-called flux tubes \citep{arturo06,yo_10,stenflo_10}. Stronger fluxes, but still two orders of 
magnitude weaker than those of the network regions, may organise at granular scales and form magnetic loops 
\citep[e.g.][]{yo_07,centeno_07,gomory_10,ishikawa_10} that appear intermittently on the quiet Sun \citep{yo_12} and have energetic 
implications on higher layers \citep{yo_10_3d}. Small-scale magnetic flux tubes have also been inferred with high-precision spectropolarimetry and high spatial resolution data \citep{lagg_10,iker_14}.

Advancing our knowledge of the very quiet Sun magnetism requires interpreting polarimetric signals in terms of 
physical quantities.  To do so, we rely on inversion codes that search for the best model parameters that fit the observed profiles. 
The model needs to be selected so that the residual of the fit and the observations is uncorrelated noise. 
This is where many works in the literature disagree and disputed results appear. An early 
controversy revolved around the strength of magnetic fields. While works in the near-infrared inferred hG fields 
\citep[e.g.][]{khomenko_03,yo_08}, other works used visible lines
and derived kG fields \citep[e.g.][]{hector_02,ita_03}. Later, the Hinode 
satellite provided spectro-polarimetric data in the visible with improved spatial resolution. In these data, hG 
were found in the internetwork \citep[e.g.][]{david_07}. Today, most of the community agrees that magnetic fields in the very quiet 
Sun are in the hG regime or even weaker. However, although one controversy was solved, the analysis 
of Hinode data has recently instigated another controversy on 
the inclination of magnetic fields: while some works point towards a mostly isotropic distribution 
\citep{yo_08_mus,lites_08,andres_09,stenflo_10}, others claim that the fields are mostly horizontal \citep[][]{david_12,david_luis_12,luis_12}. We still do not fully understand the reasons 
behind these disagreements. Some works claim that the observed profiles have not enough information to constrain model parameters \citep{yo_06}, and some others show 
that the photon noise can introduce biases and degeneracies in the inversion problem \citep{andres_09,stenflo_10,juanma_12}.

This paper presents high-precision observations with high spatial resolution of
the very quiet Sun with two Zeeman-sensitive Fe\,{\sc i} lines at 1565 nm. The
line at 1564.8 nm has a Land\'e factor of 3. This, together with its
H-band wavelength, turns it into one of the most sensitive lines to
polarisation, and one of the best lines to measure magnetic fields in the Zeeman
regime within the reach of current instrumentation. The observations  were performed with the new German GREGOR telescope,
achieving an unprecedented spatial resolution at 1.5 $\mu$m of 0.4$''$. In this work, we
analyse  these high-quality data to shed some light on the apparent
controversy surrounding the topology of the magnetic field in the
very quiet Sun.

\section{Observations}

On 2015 September 17 we recorded 1.5 $\mu$m spectro-polarimetric data of 
a quiet region at disc centre, with a spectral sampling of 40.1 m\AA. We used the Tenerife 
Infrared Polarimeter working with the the GRegor Infrared Spectrograph \citep[][]{tipII,gris} 
installed at the German GREGOR telescope \citep{gregor}. The adaptive optics system \citep{AO_gregor}
was locked on granulation and provided a spatial resolution of $\sim$0.4$''$, as inferred 
from the power spectrum of the continuum image. We scanned an area of 
$61.6''\times 13.5''$, using a time integration of 4.6 s per slit position (taking into account overheads, the 
cadence was $\sim$8s) and a step size of 0.135$''$. The data set used in this work is the 
same as in Lagg et al. (2016).

The data were corrected for flatfield and bias and were demodulated with the 
dedicated software \citep{reduccion_tip}. Additional corrections were applied to 
match the continuum of the atlas by \cite{livingston91} following a similar strategy as in 
\cite{allende04}. We modelled the average quiet-Sun intensity profile as the addition of 
the atlas convolved with a Gaussian function and a flat spectrum, whose weight is the so-called 
white-light veil. We iteratively fitted a high-order polynomial, the white-light veil, and the 
width of the Gaussian to the average quiet-Sun intensity profile. The Gaussian width was consistent 
with the spectral resolution of the instrument, and the veil was about 9\% (similar to the 
value obtained by Lagg et al. 2016). We applied the veil correction only to the 
intensity spectra, and we divided all Stokes parameters by the inferred high-order polynomial.

We then used the principal component analysis \citep{loeve55,rees03} to minimise the uncorrelated Gaussian noise. This technique consists of constructing a base of eigenvectors from the covariance matrix of the data. 
The very few first eigenvectors contain most of the correlation and the rest contain basically uncorrelated noise. Therefore, we 
can reconstruct the data with the truncated base of eigenvectors that contain useful information and remove uncorrelated noise \citep[see 
the application of this technique for denoising solar and stellar spectra in e.g.][]{yo_08, yo_08_pca}. To ensure that 
we did not lose any physical information, we reconstructed the data with 41 eigenvectors.
Unpolarised and polarised interference fringes \citep{semel03} are evident in the eigenvectors of all the 
Stokes parameters. We removed them in the truncated base of eigenvectors by fitting sinusoidal functions. Other 
artefacts, such as jumps created by the CCD separations, were also removed from the eigenvectors. After those corrections, 
we re-orthogonalised the truncated base of eigenvectors using the Gram-Schmidth algorithm. After the PCA 
denoising, the noise level in polarisation is $\sigma_n = 10^{-4}$ $\langle$I$_\mathrm{c}\rangle$, with $\langle$I$_\mathrm{c}\rangle$ the continuum intensity averaged over the field of view (FOV). The polarisation continuum was set to zero, with a precision given by the noise level.

Figure \ref{mapas} displays the intensity and polarisation maps after our denoising procedure, 
showing the high quality of the GRIS data. The circular and linear polarisation signatures 
permeate the quiet Sun:  96\% of the observed area has either linear or circular 
polarisation amplitudes above 4$\sigma_n$. Interestingly, 74\% of the 
observed area has linear (either Stokes $Q$ or $U$) polarisation and circular 
polarisation above 4$\sigma_n$. The spatial distribution of Stokes $V$ or Stokes $Q$ or $U$ are very similar in the 
sense that the stronger signals form a filamentary pattern with dead calm areas of $\sim$10$''$ diameter 
(40 Mm$^2$ area) that contain the weaker signals. 

The yellow and orange contours overplotted on the circular polarisation 
in Fig. \ref{mapas} represent the linear polarisation ($\sqrt{Q^2+U^2}$) with 
1$\times 10^{-3}$ $\langle$I$_\mathrm{c}\rangle$ and 1.5$\times 10^{-3}$ $\langle$I$_\mathrm{c}\rangle$, respectively. Most of the linear 
polarisation links circular polarisation regions with opposite polarities, but does not 
form simple $\Omega$-shaped loops. We obtained simpler structures when we adopted a higher threshold for the linear polarisation. 
Taking only the orange contours into account, we count 47 loop-like 
structures, which gives an emergence rate of 1.7 loop arcsec$^{-2}$ h$^{-1}$. This rate has been calculated as the number of 
loops divided by the scan area and the loop lifetimes. The lifetime of a loop is defined as the time in which a loop is recognised as two opposite polarities 
linked by linear polarisation. This lifetime is $\sim 2$ min and has been derived from previous works by \cite{yo_09} and \cite{yo_12}. This 
emergence rate is safely computed from raster scans when the exposure time is much shorter than the loop lifetimes. 
When the exposure time is of the order of the loops lifetime, the emergence rate is overestimated \citep[e.g. the value of 0.3 loop arcsec$^{-2}$ h$^{-1}$ 
that can be obtained using previous 1$''$ TIP-II data, as were presented in][]{yo_07}. 
The present emergence rate is almost seven times higher than the rate inferred from IMaX data 
using the Fe\,{\sc i} visible line at 525 nm at $\sim$0.15'' \citep[0.25 loop arcsec$^{-2}$ h$^{-1}$][]{yo_12}.

\section{Model approaches to infer the physical parameters}

A critical point in any inference problem is selecting the 
parametric model that is used to reproduce the observations. 
In our data, about 48\% of Stokes $V$ profiles are not regular
and have only one or more than two lobes. 
Stokes $V$ signals also exhibit area and amplitude asymmetries. 
In many cases, the linear and circular Stokes profiles are not
mutually compatible. The line-of-sight (LOS) velocity inferred from Stokes $Q$ and $U$ profiles is different from that
inferred from Stokes $V$. This incompatibility also occurs with regard to the width of the lines. Furthermore, there are
instances in which the Stokes parameters cannot be reproduced with a single magnetic field vector. All together, 
these are indicators that both the magnetic field and the LOS velocity show gradients within the resolution element, that is, along the optical depth 
and/or across the surface. Given the richness and complexity of the data, we interpret them in terms of 
three different models. We invert our spectro-polarimetric data using the Stokes Inversion based on Response functions code \citep[][]{sir}. 

We first adopted a model in which the resolution element is shared by a magnetic and a non-magnetic atmosphere. The non-magnetic component was assumed to contain the contribution of instrumental unpolarised stray light as well as 
possible magnetic substructure, that is, mixed polarities or extremely weak fields that have no net contribution to the polarised spectrum. The magnetic atmosphere was assumed to have a constant magnetic field vector and velocity along the LOS. The temperature stratification in both atmospheres was modified starting from the original VAL-C \citep{vernazza_81} model with a maximum of five nodes, forcing the temperature of the continuum to 
be the same in both components. The microturbulence, macroturbulent velocity, and the LOS velocity are free height-independent parameters in both components. The filling factor $\alpha$, that is, 
the fraction of the resolution element occupied by the magnetic atmosphere, is also a free parameter. 
The total number of free parameters in this model is 18. This model allowed us to reproduce the general properties of two-lobed Stokes $V$ profiles with low area and amplitude asymmetries and with 
compatible linear polarisation profiles. It also allows the comparison with previous studies with high Zeeman sensitivity. In the following, 
we refer to this model as model 1C.

Secondly, we modelled the Stokes profiles with two magnetic atmospheres and a global, unpolarised stray-light component (model 2C). The magnetic 
field vector of the two components was assumed constant, as well as the LOS velocity. The temperature stratifications in the two components were 
variable and had a maximum of five nodes. {We only constrained the temperature of the two components at the continuum to be the same.} 
In this model, the global stray-light profile is 
a synthetic profile (to avoid correlation between the observed profiles through fringes and other systematics) obtained as the best fit to the average Stokes $I$ in the FOV. The filling factor of the stray-light component 
was fixed to 30\% and was estimated from the absorption of the $\pi$ component in the umbral core of sunspot observations (see Borrero et al. 2016, this volume). 
Since this value is approximate, we checked that a percentage of the stray-light factor from 20\% to 60\% did not significantly change the results. The model has 21 free parameters. We are aware that some of the complexity in the Stokes profiles that this model attempts to reproduce may be produced by local polarised stray light. We do not exactly know the level of this stray light, and since we do not study particular patches but the statistics in the FOV, we did not attempt to model it. A list of the free parameters used in models 1C and 2C can be found in Table \ref{tabla}.

\begin{table*}[h!]
\centering
\caption{Free parameters used in models 1C and 2C. The superscripts 1 and 2 represent the first and second component. The variables are the temperature $T$, the LOS velocity $v_{LOS}$, the strength $B$, inclination $\theta$, and azimuth $\chi$ of the magnetic field, the macroturbulent velocity $v_{mac}$, the microturbulent velocity $v_{mic}$, the filling factor $\alpha$, and the fraction of stray light $\alpha_{SL}$.} \label{tabla}
\begin{tabular}{ccccccccccccccccc}
  & $T^1$ & $v_{LOS}^1$ & $B^1$ & $\theta^1$ & $\phi^1$ & $v_{mac}^1$ & $v_{mic}^1$ & $T^2$ & $v_{LOS}^2$ & $B^2$ & $\theta^2$ & $\phi^2$ & $v_{mac}^2$ & $v_{mic}^2$  & $\alpha$ & $\alpha_{SL}$\\
\hline
1C & 5 & 1 & - & - & - & 1 & 1 & 5\tablefootmark{1} & 1 & 1 & 1 & 1 & 1\tablefootmark{2} & 1 & 1 & - \\
2C & 5 & 1 & 1 & 1 & 1 & 1 & 1 & 5\tablefootmark{1} & 1 & 1 & 1 & 1 & 1\tablefootmark{2} & 1 & 1 & 30\%\\
\hline
\end{tabular}
\tablefoot{
\tablefoottext{1}{The temperature at the continuum of the second component is forced to be the same as in the first component.}
\tablefoottext{2}{The macroturbulent velocity of the second component is forced to be the same as the first component.}
}
\end{table*}

\section{One magnetic structure embedded in a field-free atmosphere}

To have a general view of the quiet-Sun magnetism, Fig. \ref{mapas_inv} displays the maps of the inferred magnetic field strength $B$, the magnetic field inclination $\theta$, 
the magnetic field azimuth $\chi$, and the filling 
factor using a single height-independent magnetic atmosphere (model 1C). Most of the area is full of weak hG fields, although there are a few patches with kG fields (white contours). 
These kG patches coincide with vertical fields, but only the largest patch (at [20,1.5] Mm) 
has filling factors higher than 15\%. In general, vertical 
fields correspond to stronger field strengths,
while the more horizontal fields have strengths of $\sim$150 G. This behaviour is the same as that observed
by \cite{yo_08} with the same instrument but with lower spatial resolution. 

In 49\% of the FOV, model 2C yields a fit of similar quality as model 1C for Stokes $Q$, $U$, and $V$, simultaneously. In this case, we prefer the model with fewer free parameters. From these pixels, we selected those with amplitudes larger than 4$\sigma_n$, with 
two-lobed Stokes $V$ profiles. For these pixels, which account for 23\% of the FOV, the magnetic flux density $\phi = B\alpha\cos{\theta}$ is well constrained by the observed profiles. The 
magnetic flux density is defined through the solar surface, in our case, the normal to the surface coincides with the LOS. We
note that at the disc centre, the magnetic flux density only depends on the 
LOS magnetic field, that is, on the Stokes $V$ profile.

We performed several inversions at each pixel in which we fixed the inclination to values from 5 to 85 deg and let the code find a solution with the field strength, the filling factor, and the microturbulent velocity as free variables. We defined the normalised rms of the polarisation profiles for each inversion with a fixed value of the inclination as 
\begin{equation}
\sum_{S=QUV}\frac{\sigma(S^{\mathrm{fit}}-S^{\mathrm{obs}})}{3\sigma_n}.
\end{equation}
The symbol $\sigma$ stands for the standard 
deviation operation, $S^{\mathrm{obs}}$ for the 
observed polarisation profiles, and $S^{\mathrm{fit}}$ for its best fit. When the normalised rms in these inversions varies by more than 20\% (and the individual values are below 1.2, i.e., reasonable fits), we considered 
the magnetic field vector to be well constrained by the observations, since the inversion code is not able to reproduce the observations equally well with different inclination values. The pixels with regular Stokes $V$ profiles,  where the magnetic field is 
reliably recovered using a single atmosphere approach, represent 12\% of the FOV. 

Figure \ref{hist_flujo} shows the histograms of the magnetic flux density for the pixels in which the magnetic field 
vector is not reliably inferred (solid line) and for those where it is well constrained by the observations (dotted line). 
We find stronger fluxes for the latter pixels. 

The histograms of the 
magnetic field vector and the filling factor of the pixels where they are well constrained by the observations 
are displayed in Fig. \ref{hist_inv}. Most of the fields are of the order of hG: 80\% of the pixels have 
strengths below 450 G and 50\% are below 250 G. The field strength
distribution has a peak 
at $\sim$250 G. These results for the field strength are compatible with previous studies using near-infrared data at lower resolution 
and longer exposure times \citep{khomenko_03,yo_08}. However, the peak at $\sim$250 G was located at $\sim$450 G in these studies. Lowering the spatial resolution or increasing the exposure time 
dilutes the spectro-polarimetric signals and results in a higher detection limit for the field strength. 

The inclination distribution (top right panel in Fig. \ref{hist_inv}) has
an excess of horizontal ($70^\circ-110^\circ$) fields with respect to an isotropic
distribution (that follows $p(\theta) = \sin{\theta}$ and $p(\chi) = 1$). Surprisingly, the flux in the two polarities is not balanced in the observed FOV.
Previous  studies of very quiet regions reported a polarity balance, even with
smaller observed areas \cite[e.g.][]{lites_02,ita_03,yo_08}. This imbalance
occurs for all signals, hence it cannot be a result of unipolar
network patches or a unipolar weak component previously undetected.

\begin{figure}
\centering
\includegraphics[viewport=0 0 410 350,width=\columnwidth]{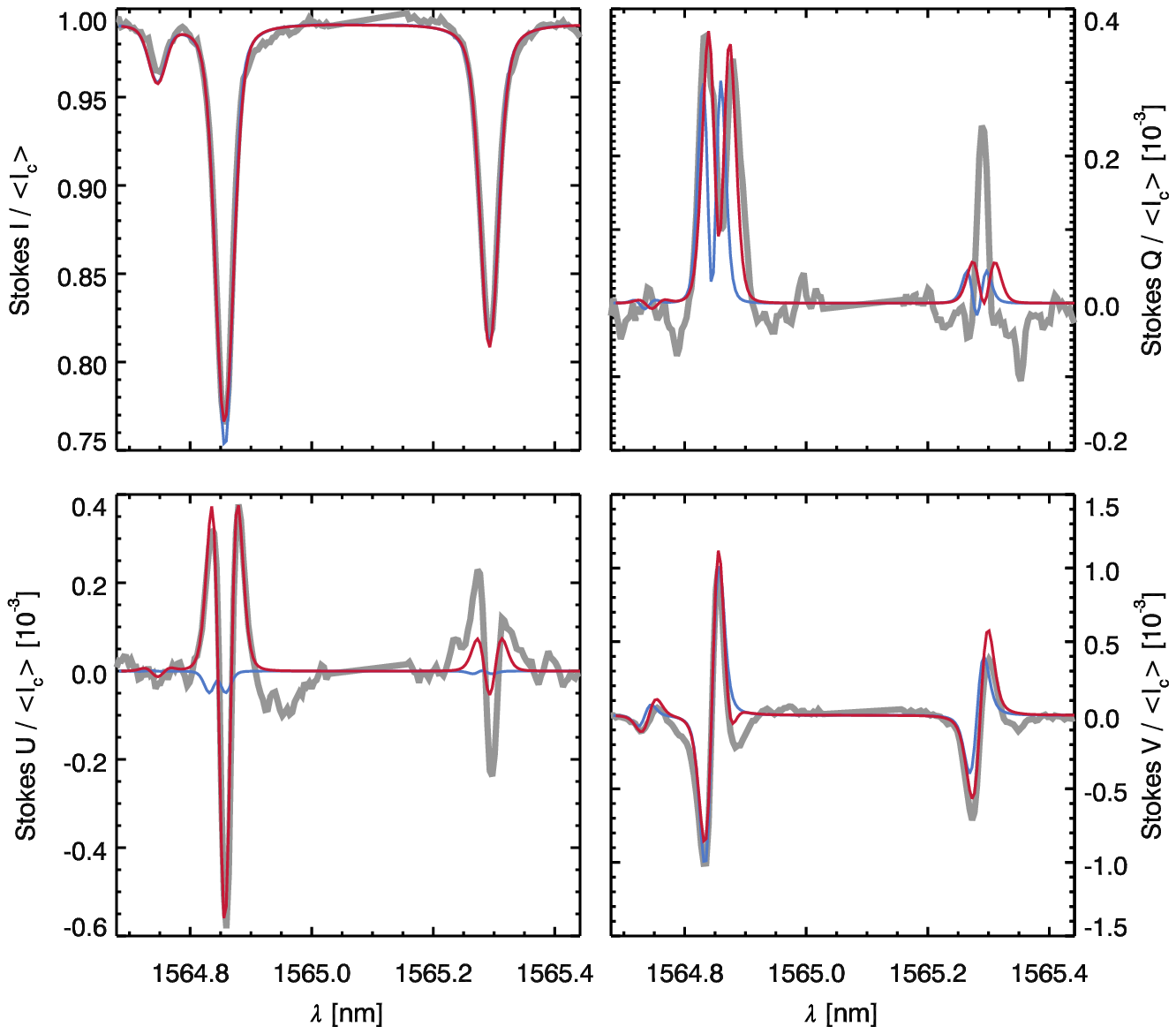}
\includegraphics[viewport=0 0 410 350,width=\columnwidth]{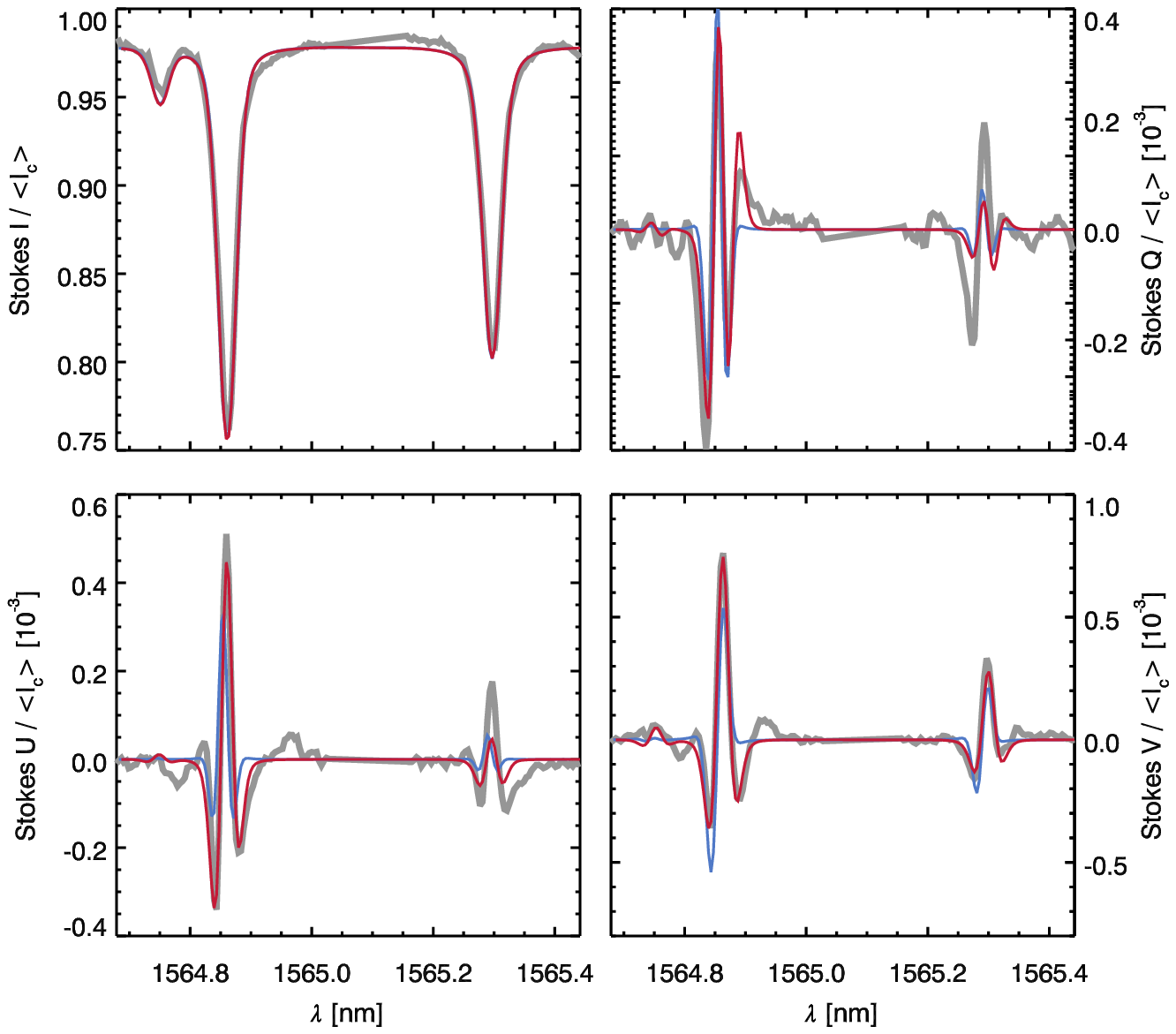}
\includegraphics[viewport=0 0 410 350,width=\columnwidth]{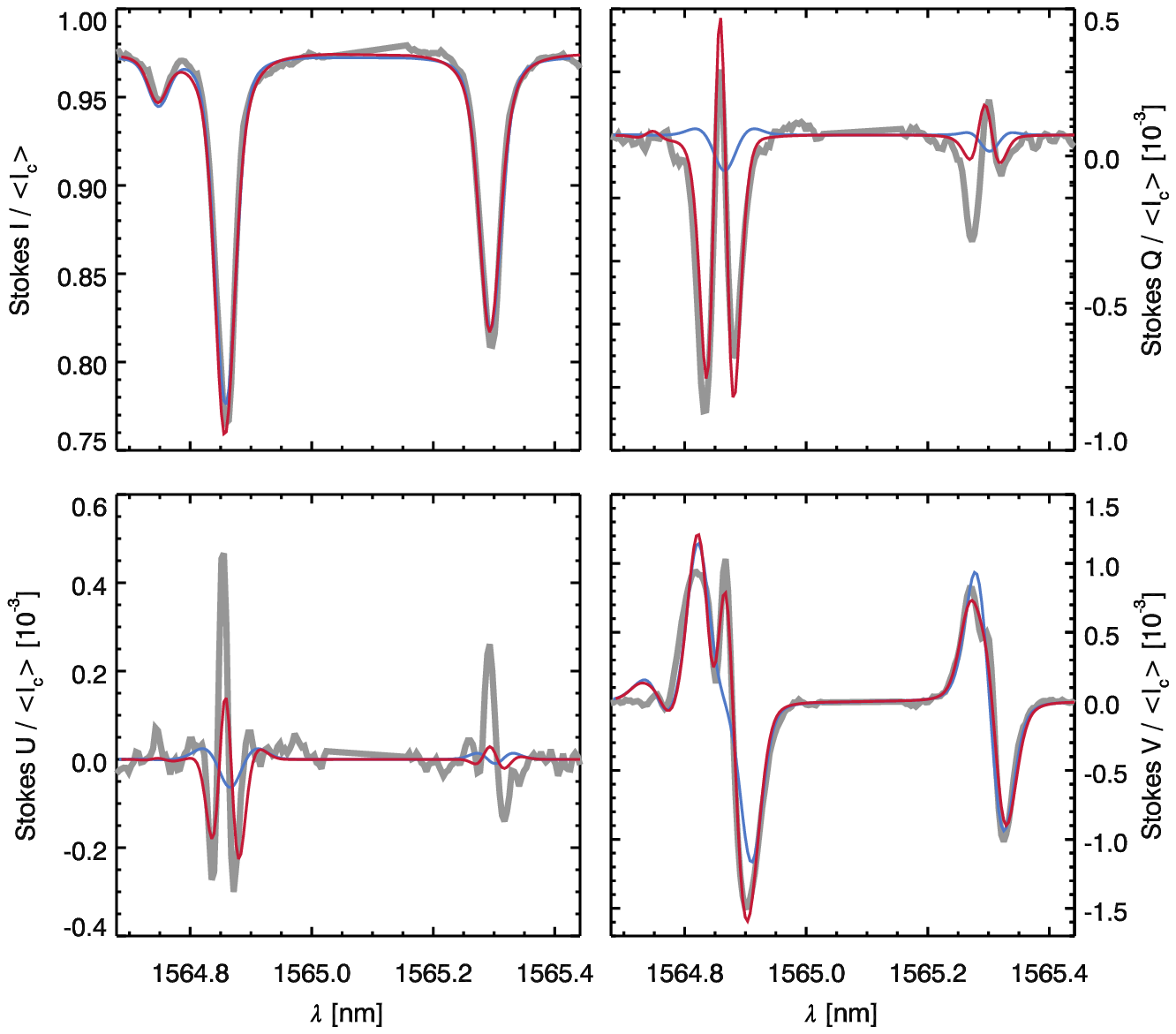}
\caption{Three typical examples of Stokes profiles, displayed as the thick grey line. The blue line represents the inversion 
with the model that considers only one magnetic component embedded in a non-magnetised atmosphere. The 
red line displays the inversion using a model with two magnetic atmospheres coexisting with the 30\% 
contamination of unpolarised stray light.}
\label{perf_mag+nm_2mag}
\end{figure}

\begin{figure*}[!t]
\centering
\includegraphics[width=0.4\textwidth,viewport=43 9 493 336]{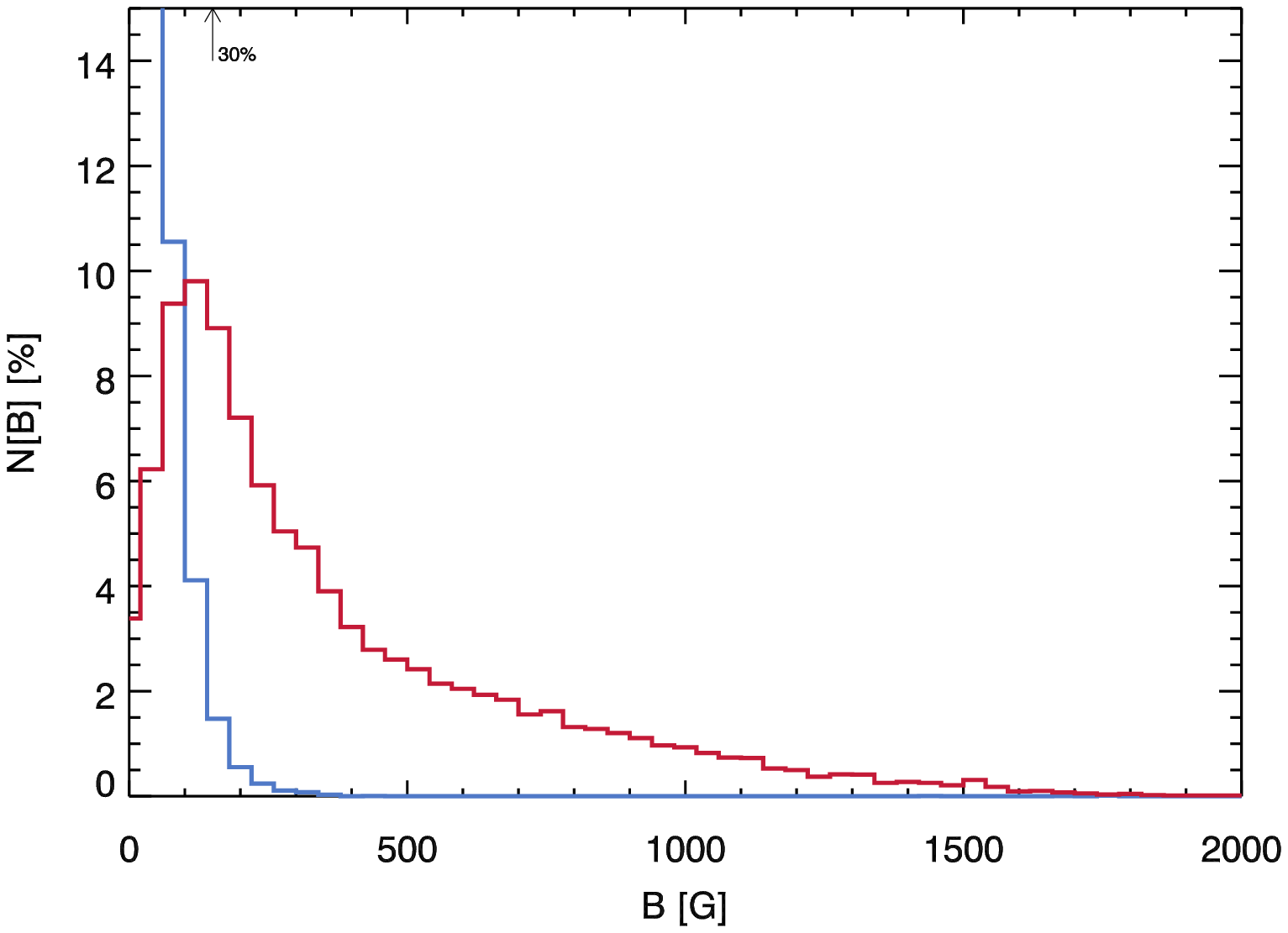}
\includegraphics[width=0.4\textwidth,viewport=43 9 493 336]{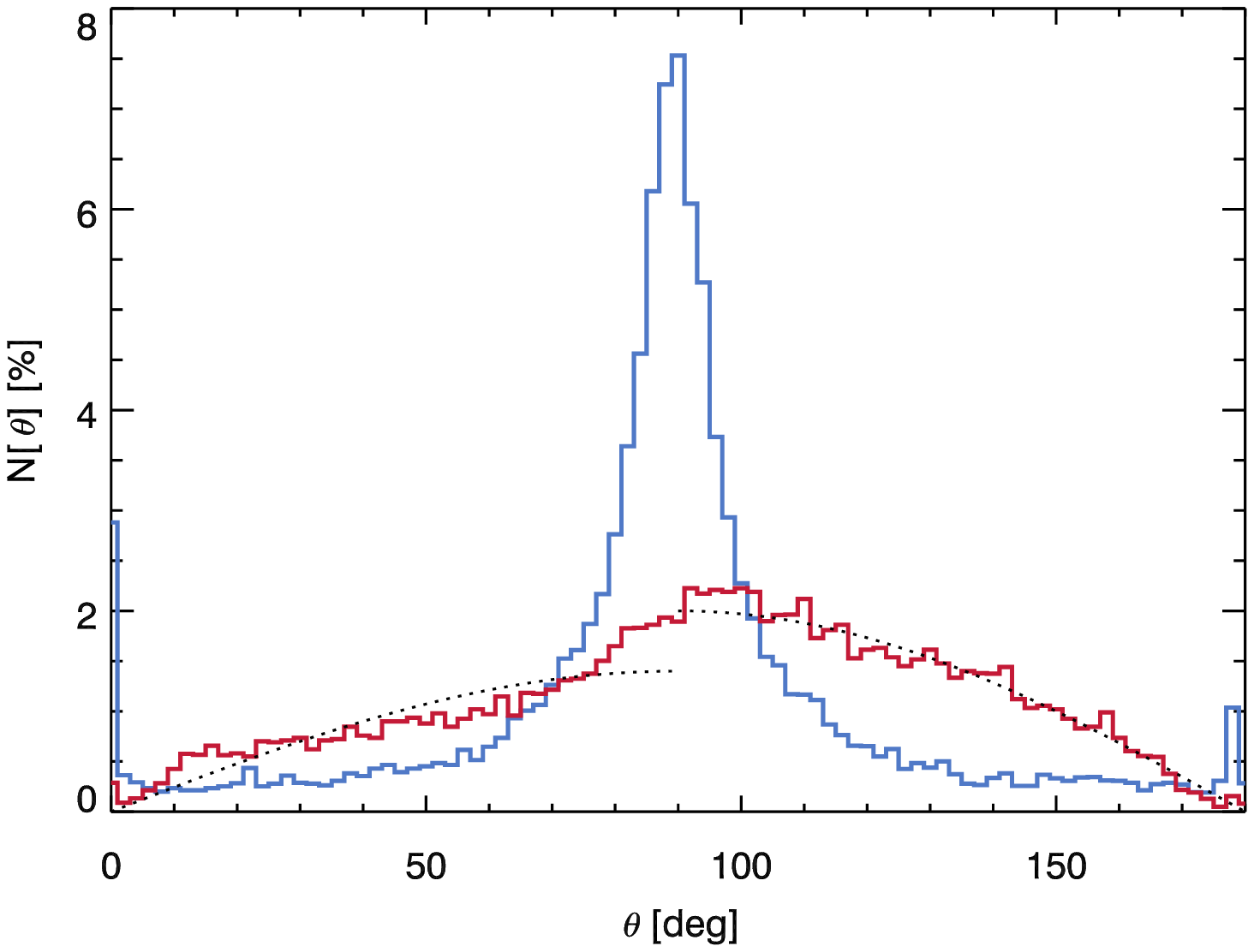}\\
\includegraphics[width=0.4\textwidth,viewport=43 9 493 336]{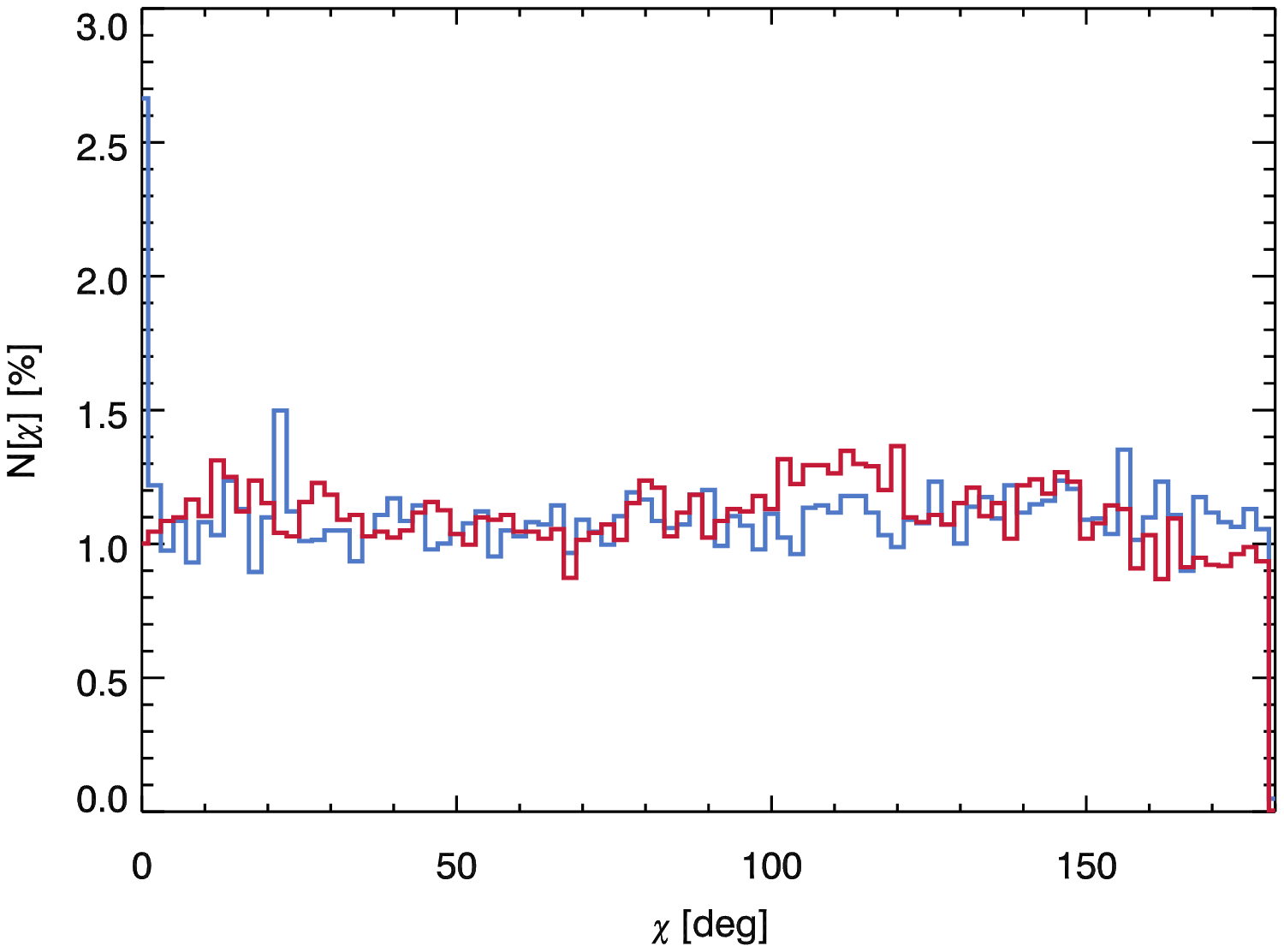}
\includegraphics[width=0.4\textwidth,viewport=43 9 493 336]{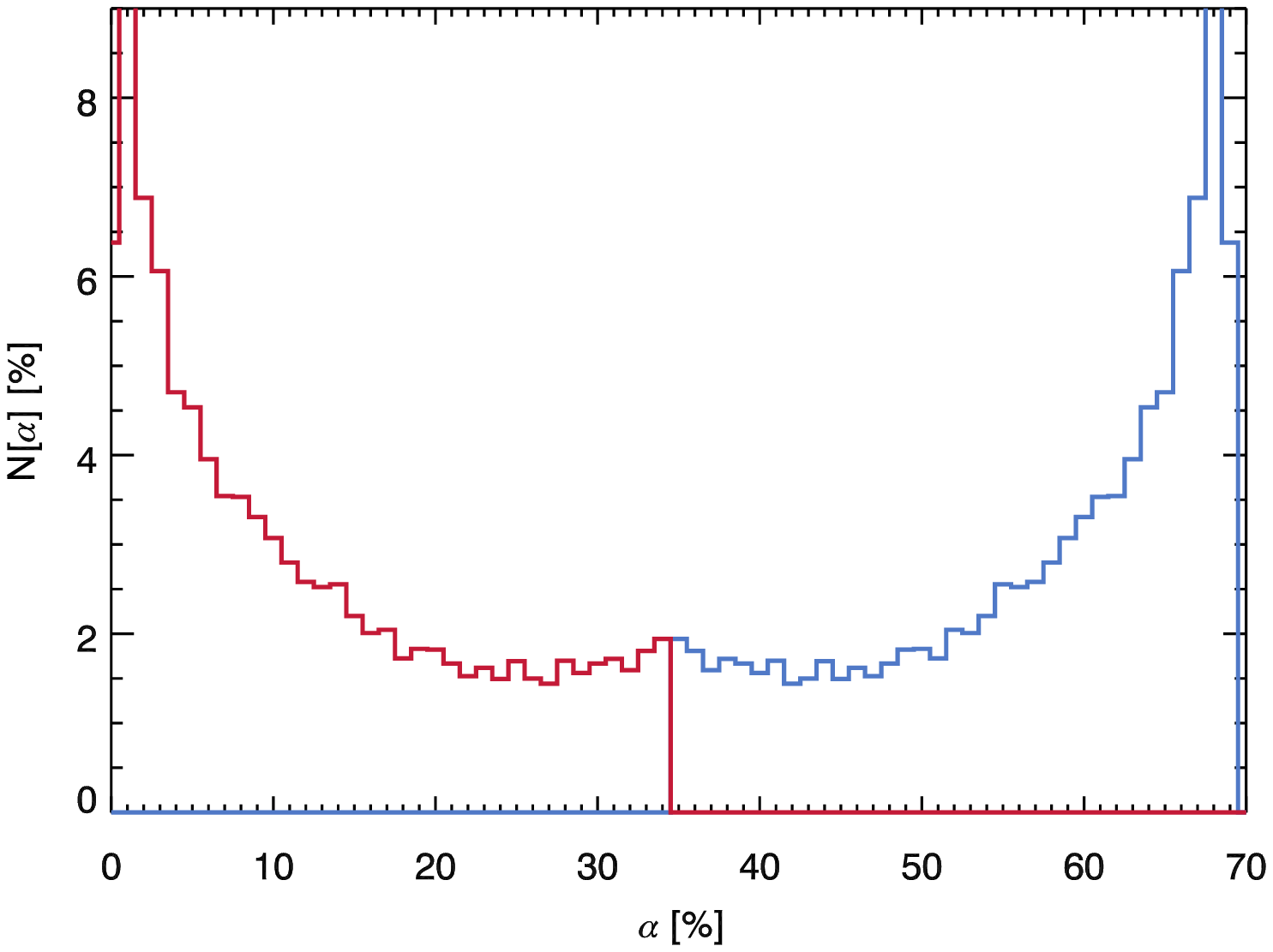}
\caption{Histograms of the inferred field vector (strength, inclination, and azimuth) and the filling factors 
using a model with two magnetic atmospheres contaminated by 30\% of 
unpolarised stray light. All the inverted pixels have Stokes $Q$ or $U$ or $V$ signals above 4$\sigma_n$ and have better 
fits than the single magnetic component inversion. 
The blue (red) lines represent the large (small) component, the one with larger (smaller) filling factor.
The dotted grey lines superposed to the inclination histogram represent a $\sin{\theta}$ 
distribution. Taking into account that the azimuth is uniform, these dotted lines represent an isotropic distribution of the 
field vector.}
\label{hist_inv_2mag}
\end{figure*}

\begin{figure}[!t]
\centering
\includegraphics[width=0.9\columnwidth,viewport=20 8 489 340]{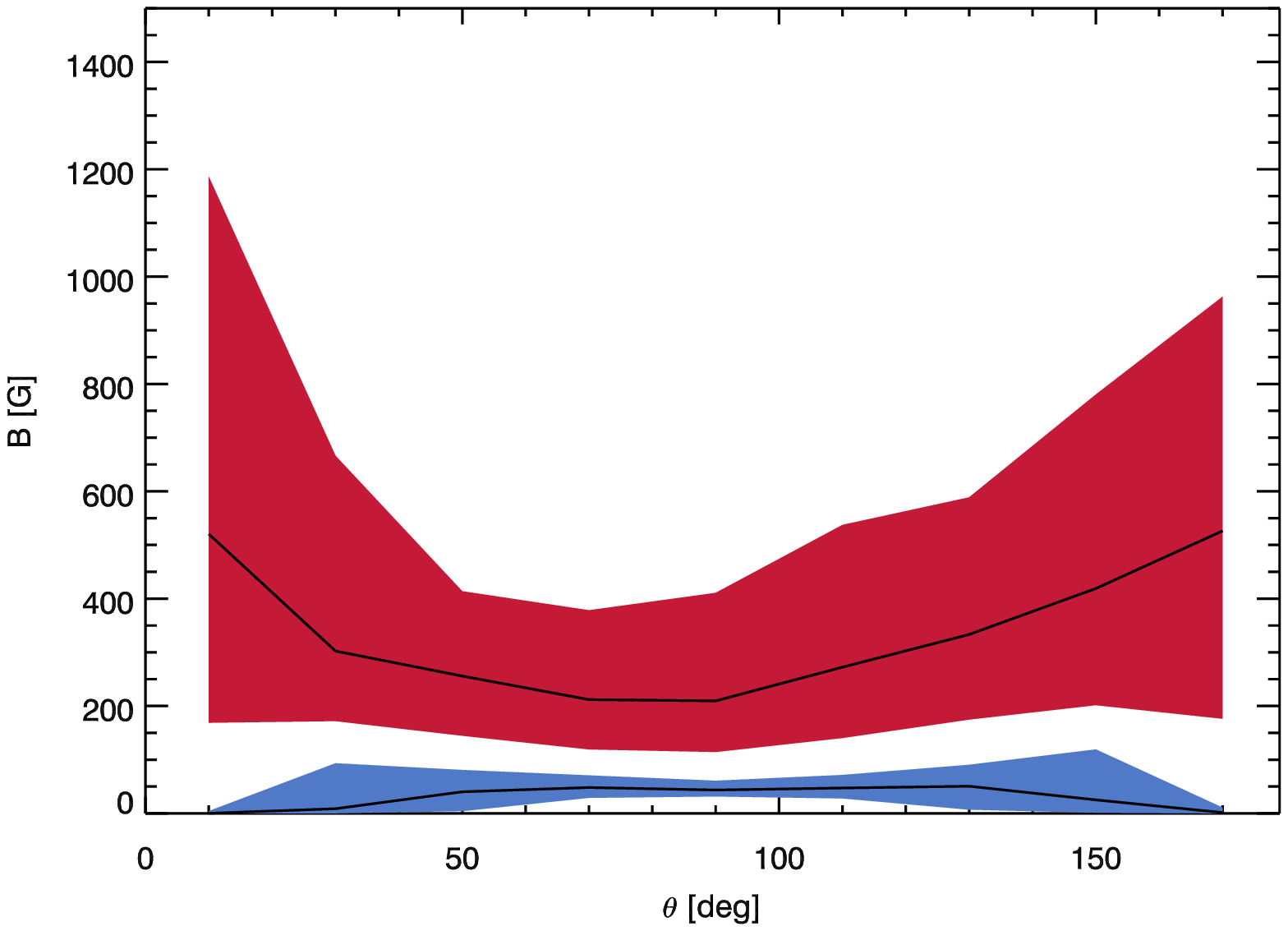}
\includegraphics[width=0.9\columnwidth,viewport=20 8 489 340]{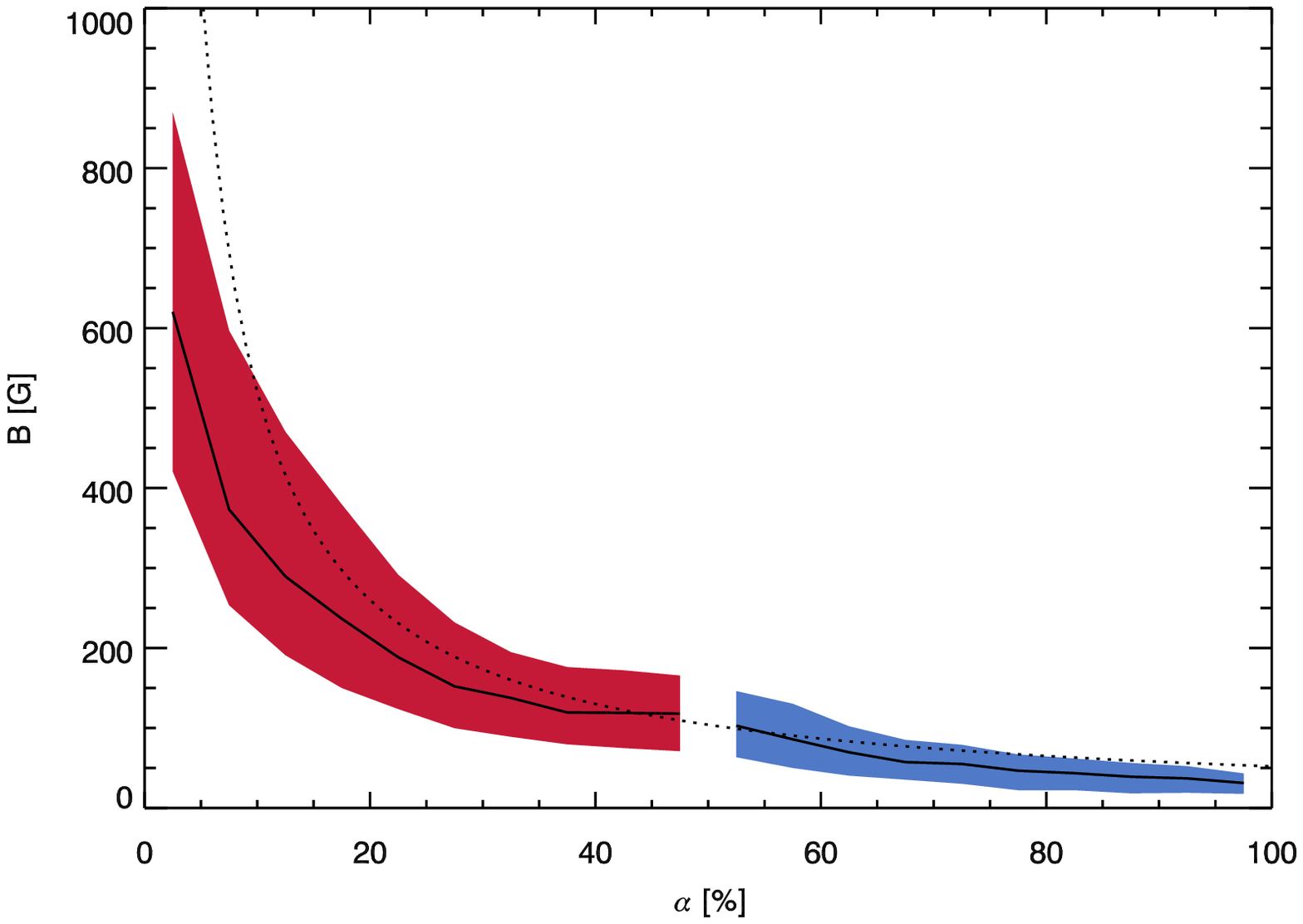}
\caption{Magnetic field strength versus inclination (top) and versus filling factor (bottom). 
All the inverted pixels have Stokes $Q$ or $U$ or $V$ signals above 4$\sigma_n$. Blue (red) 
represents the large (small) component, the one with larger (smaller) filling factor. 
The black line represents the median value, and the shaded areas enclose percentiles 25 to 75. 
The dotted black line is an hyperbolic relationship given by 5200/$\alpha$[\%].}
\label{box_inv_2mag}
\end{figure}

\begin{figure}
\centering
\includegraphics[viewport=8 0 387 350, width=\columnwidth]{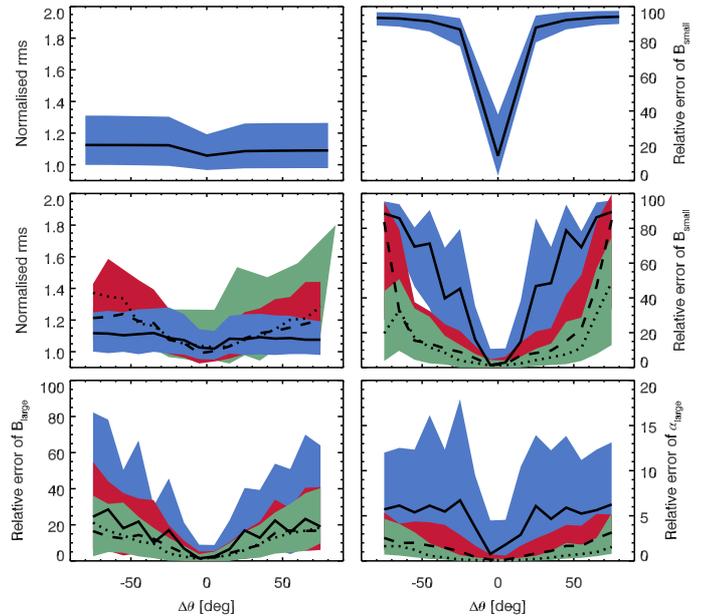}
\caption{Top panels: results of the numerical test to check the validity of the inferred magnetic properties of the 
large component. The top left panel represents the the normalised rms of the polarisation profiles defined as 
$\sum_{S=QUV}\sigma(S^{\mathrm{fit}}-S^{\mathrm{obs}})/3\sigma_n$ plotted versus the inclination we have forced in each inversion. The top right panel 
shows the relative modification of the magnetic field strength of the large component to account for the 
forced values of the inclination. The 
black line represents the 50 percentile (median value), and the shaded blue areas enclose percentiles 25 to 75. The remaining panels show the numerical test to check the reliability of the small component. The middle left panel represents the 
normalised rms of the polarisation profiles for the weak population (solid line for the median and blue shaded areas for percentiles 25 
to 75), the strong population (dashed line and red shaded areas), and the very strong population (dotted line and green shaded areas). 
The middle and bottom right panels 
show the relative modification of the magnetic field strength of the small component and the filling factor to account for the 
forced values of the inclination.}
\label{test_cdebil}
\end{figure}

\begin{figure*}
\centering
\includegraphics[width=0.8\textwidth]{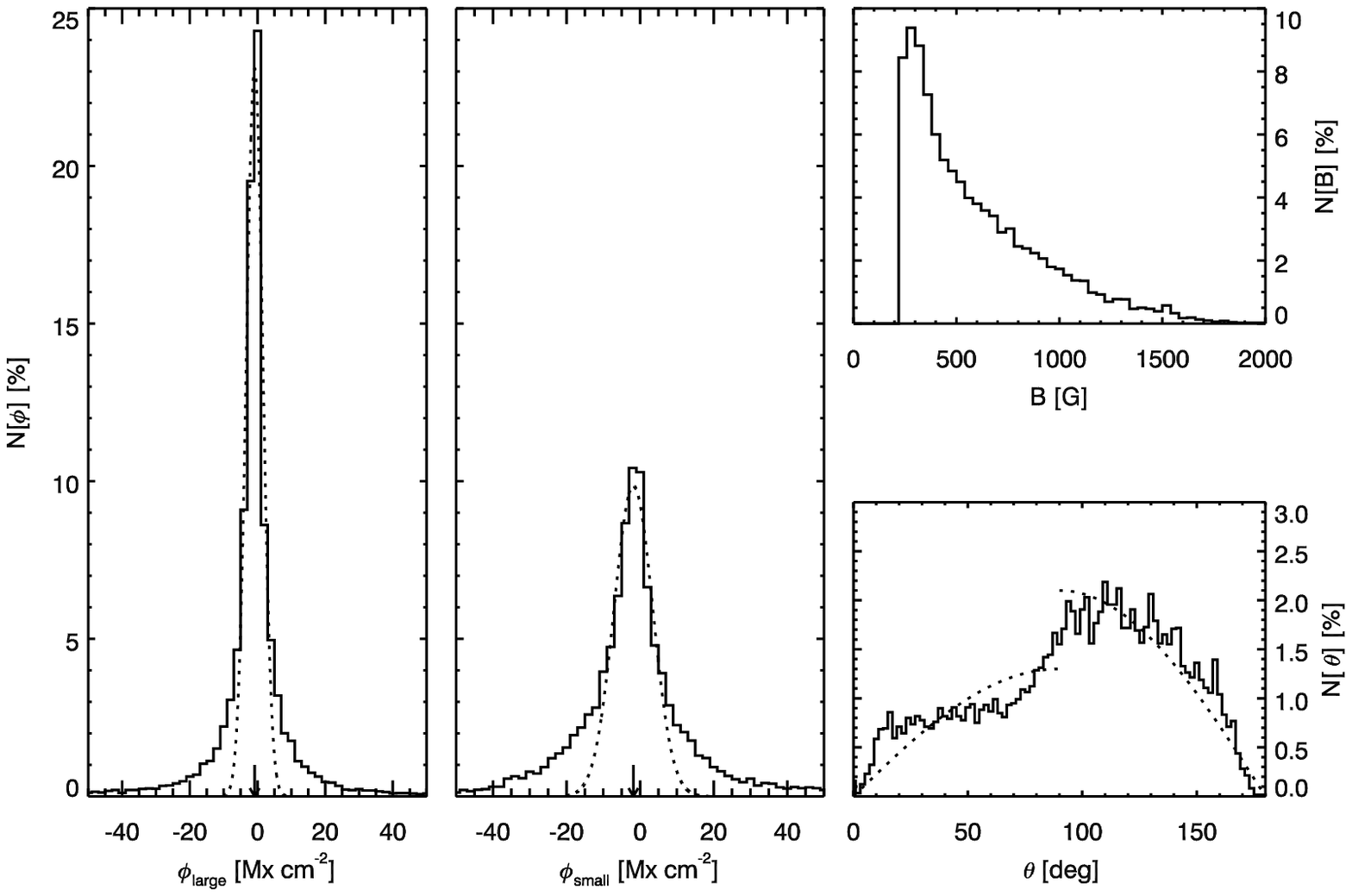}
\caption{Left panel: histogram of the magnetic flux density inferred for the large component. Middle panel: histogram of the 
magnetic flux density of the small component for those pixels with fields strengths below 250 G. 
The vertical arrow marks the average value of the fitted Gaussian core (dotted line). The top and bottom right panels 
display the magnetic field strength and inclination of the small component for field strengths above 250 G.}
\label{flux_large_component}
\end{figure*}

\section{Two magnetic atmospheres and stray light}

In 51\% of the FOV, the reduced chi-square of Stokes $Q$, $U$, and $V$ of the inversions with model 2C was smaller than in model 1C inversions. Figure \ref{perf_mag+nm_2mag} shows three examples in which the fit with two magnetic components is better 
than the fit from a single magnetic component and a non-magnetic atmosphere. In the top panel, the Stokes $V$ profile is 
well fitted by the two models, but for Stokes $Q$ and $U$ the single magnetic component 
fails to reproduce the amplitudes and their Doppler shift. The middle and bottom panels represent irregular Stokes $V$ profiles 
that cannot be fitted with a single magnetic field; they require a more complex model with two magnetic components. 

The results of the inversions for those pixels that are fitted better by model 2C are presented in Fig. \ref{hist_inv_2mag}. The large (small) component is defined as the one with 
the larger (lower) filling factor. The two components have clearly different magnetic properties. The 
larger component has weak field strengths, 80\% are 
below 90 G and 50\% are below 50 G, the inclination
distribution has a peak at 90$^\circ$, and the azimuth has a uniform distribution. 
The small component has stronger fields, for which 80\% are below 750 G and 50\% are below 350 G, and it is almost isotropic, with a 
slight polarity imbalance. 

Figure \ref{box_inv_2mag} displays the inferred inclination and
filling factor values versus the magnetic field strength. The  large component
(shown in blue) has an almost constant field strength for inclinations between 50$^\circ$ and 130$^\circ$. 
The more vertical fields (around 30$^\circ$ and 150$^\circ$) have slightly stronger,
but still very weak field strengths ($\sim$100-150 G on average). The more horizontal fields of the small component have the weakest fields, 
and the more vertical are stronger,
reaching values above 1 kG. The relationship
of the filling factor and the field strength is
compatible with a hyperbola given by $B \approx 5200/\alpha$[\%] (dotted line) for fields below 300 G. 
This behaviour, and the field strengths involved, led us to assume that there is not 
enough information in the polarised spectra of the weakest signals to reliably infer 
the field strength. In these cases, the only quantity that could be inferred without ambiguity 
was the magnetic flux density. Above 300 G (70\%), field strength,
inclination, and filling factor are unambiguous.

To check the reliability of the magnetic parameters inferred for the large component, we performed the 
following numerical test. We selected 1000 pixels from the sample used for Fig. \ref{hist_inv_2mag} with 
inclinations of the large component between 80 and 100 deg. We
note that half the area of the distribution has 
inclinations between 80 and 100 degrees. The remaining pixels that are not as horizontal are 
very likely in the first order weak-field regime of the Zeeman effect given their extremely low field strengths. This regime 
implies that the field strength, inclination, and filling 
factor are degenerated quantities. For the low field strengths, the only way to avoid the 
weak-field regime is to have a field inclined enough to generate linear polarisation above the noise level. For this reason, we performed the 
reliability test for the more inclined fields of the large component. 

We inverted the more horizontal pixels of the large component by
forcing the inclination from 5 deg to 85 deg (or 100 to 175 deg for the opposite polarity) 
and letting the code modify only the magnetic field strength 
and the microturbulent velocity of the large component. The atmospheric parameters of the 
small component, the percentage of stray light, and the filling factor of the large 
component were fixed. The top left panel in Fig. \ref{test_cdebil} shows the normalised rms of the Stokes profiles with respect to the 
difference between the fixed inclination and the one inferred from the 2C model. 
The flat curve in this figure indicates that the code is able to find a good fit for 
all values of the inclination differences by modifying the magnetic field strength. The 
microturbulent velocity has always reasonable values below 2 km s$^{-1}$. The variation of 
the field strength as compared to the one inferred from the 2C inversion (top right panel in Fig. \ref{test_cdebil})
can reach values close to 100\% for inclination differences near 90 deg. This means that with the present spatial resolution and polarimetric sensitivity, 
there is not enough information in the spectral lines to infer the field strength and the 
inclination of the large component independently. We have checked that the only quantity that remains the same in all 
inversions with fixed inclinations and hence is the only one that can be reliably inferred for the large component is the magnetic flux density, defined as 
$B\alpha\cos{\theta}$. The histogram of the magnetic flux density of the large component is shown in the leftmost panel of \hbox{Fig. 
\ref{flux_large_component}}. It shows more extended wings than a Gaussian, and has an 
average flux density of $-0.84\pm 0.02$ Mx cm$^{-2}$ (arrow). 

To check the reliability of the magnetic parameters inferred for the small component, we
performed the following test. We fixed the inclination of the small component from 5 to 85 deg (or 100 to 175 deg) 
and let the code vary the field strength and the microturbulent 
velocity of the small component, and the field strength, inclination, and microturbulent velocity of the 
large component, as well as the filling factor. We distinguished
between three populations of the small component: weak 
fields (below 250 G), strong fields (between 250 and 800 G), and very strong fields (above 800 G). The middle left panel 
of Fig. \ref{test_cdebil} shows the normalised rms of the Stokes profiles with respect to the 
difference between the fixed inclination and the one inferred from the 2C model. For the weak fields, the code is able 
to reproduce the Stokes profiles regardless of the inclination by varying mainly the field strength of the 
small component. The variation of the field strength of the small component with respect to the 2C inversion can be 
as huge as 100\% for the cases in which the inclination difference is largest (close to 90 deg). We note that the 
behaviour of the relative error of the magnetic field strength of the small component is very similar to the one 
for the large component in the previous test. Similarly as for the large component, this means that with the present spatial resolution and polarimetric sensitivity, 
there is not enough information in the spectral lines to infer the field strength and the 
inclination of the small component for fields below 250 G. The only quantity that remains the same in all 
inversions with fixed inclinations is again the magnetic flux density. The histogram of the magnetic flux density of the small component for fields 
below 250 G is shown in the middle panel of \hbox{Fig. 
\ref{flux_large_component}}. It shows more extended wings than a Gaussian, and contains slightly more strong fluxes than the large component. It has an 
average flux density of $-1.8\pm 0.02$ Mx cm$^{-2}$ (arrow). 

For the strong and very strong fields, the code is not able to reproduce the Stokes profiles for 
all the fixed inclinations equally well. It is only capable of obtaining similar fits when the inclination difference is smaller than 
around 30 deg. It can also be seen that the relative error of the magnetic field strength of the small component is 
almost constant around 10\%. This means that the Zeeman splitting is already important, or that the linear polarisation is well above the 
noise level, or both, preventing the code from finding a good fit by varying the field strength 
if we fix an inclination different than the one inferred from the 2C inversion. Therefore, 
we conclude that the magnetic field vector and filling factor of the small component are 
constrained by the observations for fields above 250 G. The leftmost panels of Fig. \ref{flux_large_component} display the 
magnetic field strength and inclination of the small component that are reliably inferred. The results are 
very similar to those in Fig. \ref{hist_inv_2mag}. The azimuth and the filling factor are not displayed, but they are 
almost equal to those in Fig. \ref{hist_inv_2mag}. The field strengths have a distribution whose decay resembles 
an exponential function with an average field of 390 G. The inclination still shows a clear polarity imbalance. Together 
with the uniform azimuth, the topology is close to isotropic. As inferred from the validity test, the inclination has an uncertainty of about 30 deg, the 
field strength about 10\%, and the filling factor about 1\%.

Asymmetries and single or additional lobes in Stokes $V$, or the relative velocities between Stokes $V$ 
and the linear polarisation profiles, can also be explained by gradients along the optical depth. To check 
if the observations can discern vertical and horizontal gradients, we performed an inversion with a magnetic component 
(with a filling factor of 70\%) with allowed gradients of all the physical properties along the optical depth, 
except for the microturbulent velocity (model G). The remaining 30\% is filled with the same unpolarised stray-light component used in the 
2C inversions. Allowing gradients of the physical quantities along the LOS is the only way to reproduce the area asymmetry in Stokes $V$ \citep{solanki93}. 

Inversions with model G are difficult to be performed automatically to the whole map because different 
observed profiles require a different number of nodes in the stratification of the physical quantities. To check if 
the observations are better reproduced by vertical or horizontal gradients, we selected some representative profiles. To invert these selected 
examples, we proceeded by fixing the number of nodes of the temperature at a maximum of 5. For the remaining physical quantities, 
we first tried with two nodes (a linear gradient) in magnetic field strength and LOS velocity, and then increased the complexity 
until we found a good fit. The number of free parameters needed to fit the representative profiles was between 20 
(only linear gradients) up to 42 when more complicated gradients were allowed. We are aware that such large number of free 
parameters cannot be constrained by the observations and multiple solutions to the inverse problem can exist \citep{andres_12}. But this study 
is beyond the scope of this paper. Our aim here is not to extract real stratifications of the very quiet atmosphere but to see how observations are reproduced with this model G as compared to model 2C.

\begin{figure*}
\centering
\includegraphics[viewport=0 0 410 350,width=0.97\columnwidth]{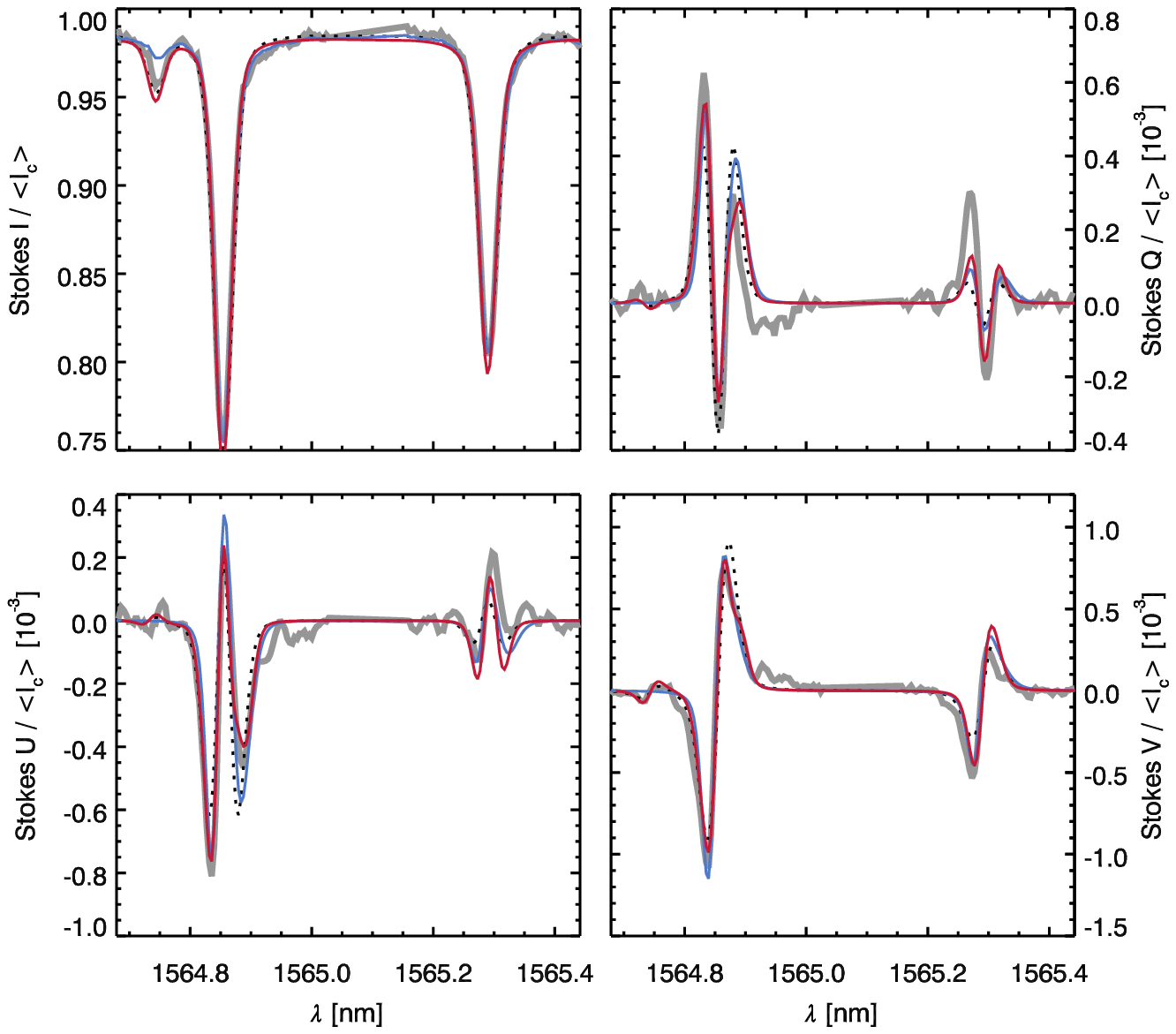}\hspace{0.5cm}
\includegraphics[viewport=0 0 410 350,width=0.97\columnwidth]{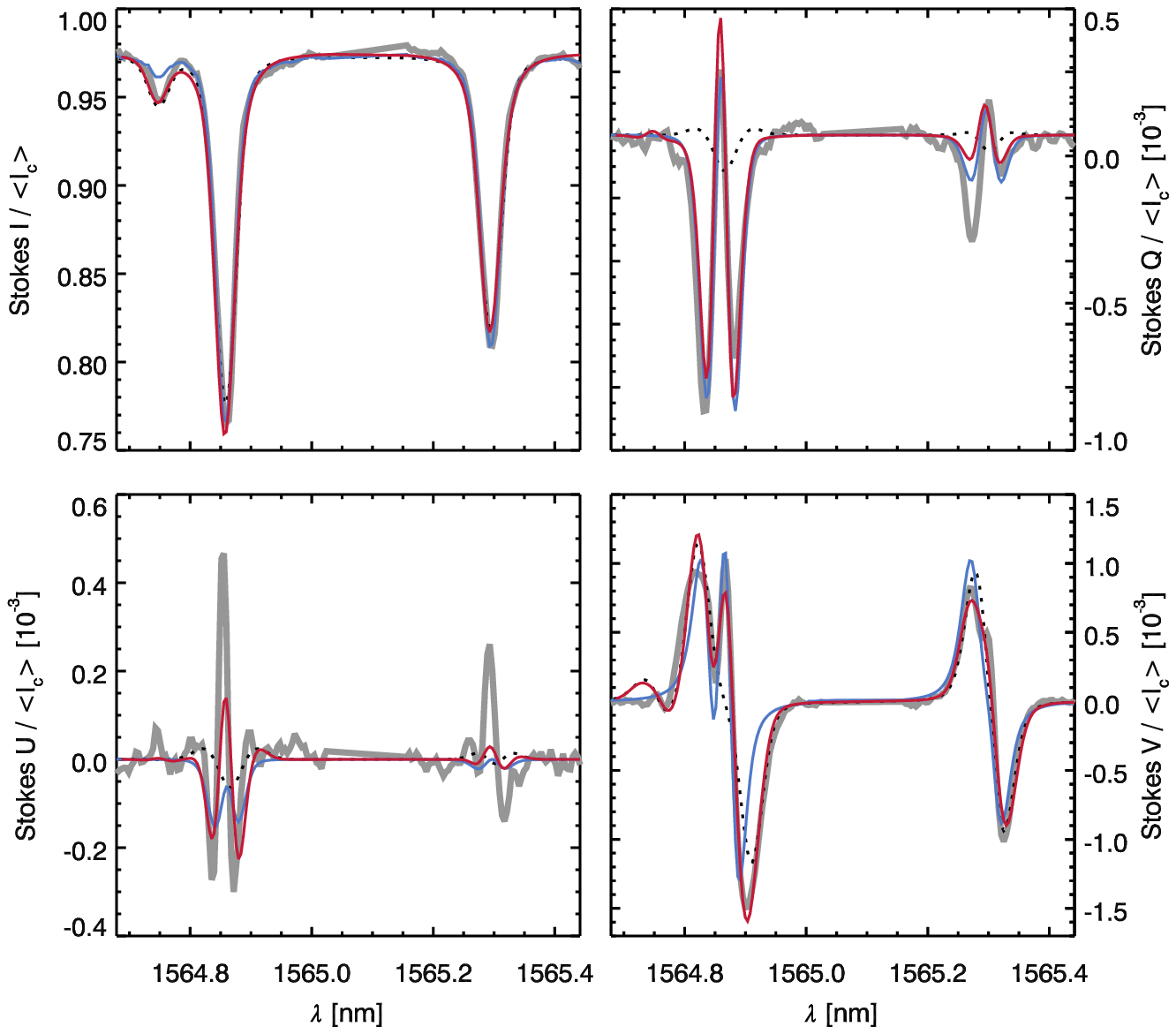}
\includegraphics[viewport=0 0 410 350,width=0.97\columnwidth]{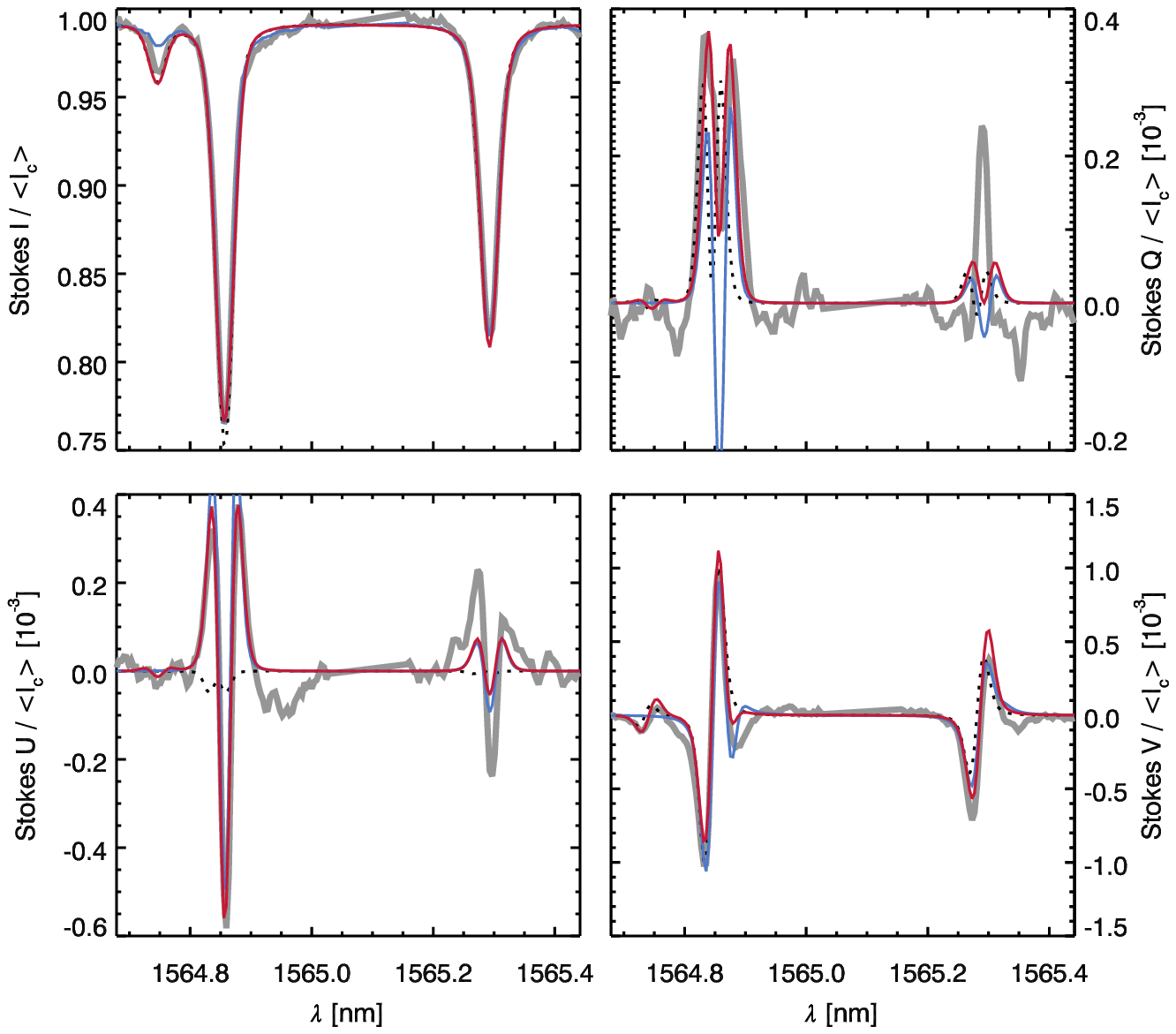}\hspace{0.5cm}
\includegraphics[viewport=0 0 410 350,width=0.97\columnwidth]{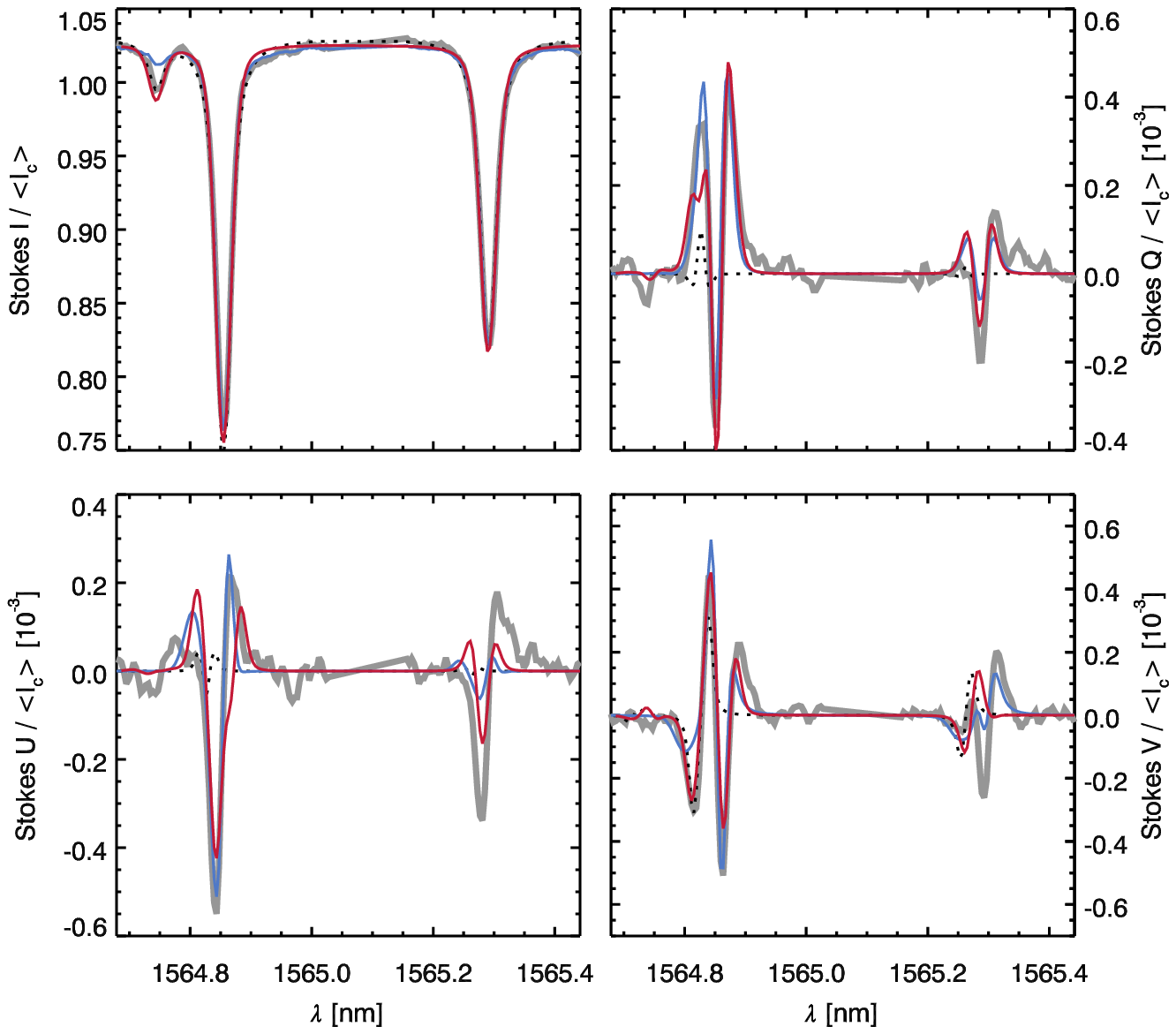}
\caption{Four typical examples of Stokes profiles, displayed as the thick grey line. The blue line represents the inversion 
using a model with two magnetic atmospheres (with constant field vector and LOS velocity) coexisting with the 30\% 
contamination of unpolarised stray light. The red line display the inversion of one magnetic atmosphere 
coexisting with the 30\% 
contamination of unpolarised stray light with allowed gradients in all physical quantities (except for the microturbulent velocity). The dotted black line represents the inversion 
with the model that considers only one magnetic component (with constant field vector and LOS velocity) embedded in a non-magnetised atmosphere.}
\label{perf_2mag+grad}
\end{figure*}

% \begin{figure*}
% \centering
% \includegraphics[viewport=0 0 410 350,width=0.97\columnwidth]{ej3_2mag+grad_modelos.eps}
% \includegraphics[viewport=0 0 410 350,width=0.97\columnwidth]{ej2_2mag+grad_modelos.eps}
% \includegraphics[viewport=0 0 410 350,width=0.97\columnwidth]{ej6_2mag+grad_modelos.eps}
% \includegraphics[viewport=0 0 410 350,width=0.97\columnwidth]{ej4_2mag+grad_modelos.eps}
% \caption{Atmospheric stratifications of the temperature, the strength and inclination of the magnetic field, and the LOS 
% velocity for the four inverted profiles shown in Fig. \ref{perf_2mag+grad}. The black line represents the model atmosphere retrieved from the inversion allowing gradients along the LOS. The blue shadowed areas are computed using the response functions and represent 
% the sensitivity of the observed Stokes profiles to each quantity given the inferred model. A smaller dispersion points to a larger sensitivity (what is usually called the "formation region" of the spectral line). Orange (red) colours represent the 
% values of the parameters obtained with the two constant magnetic component model for the large (small) component.}
% \label{perf_2mag+grad_modelos}
% \end{figure*}

Figure \ref{perf_2mag+grad} displays the three selected representative profiles to perform the inversions with model G. We chose profiles that 
have some features that show the need for a more complex scenario than a single magnetic field in the 
resolution element. The profiles 
in the top left panel are representative of those pixels with regular polarisation signatures (Stokes $Q$ and $U$ also have the same Doppler shift as Stokes $V$) with area and amplitude asymmetries. In this case, we obtain a similar fit with model G (red line) and model 2C. Moreover, 
it is not surprising that a good fit was also obtained with just one constant magnetic component (model 1C; dotted black line), since the asymmetries are not large. 

The top right and bottom panels show clear examples in which Stokes $Q$ and $U$ are not compatible with Stokes $V$. The top right panel 
shows a better fit in Stokes $V$ for model G than for model 2C. However, similar fits for model G and 2C are obtained in the bottom 
panels. In these types of profiles, it is clear that the observations are better fitted with magnetic substructure within the resolution element than with a constant magnetic field (model 1C). However, since the area asymmetries are not very large, the observations do not 
have the information necessary to distinguish between gradients along the LOS (model G) or across the surface (model 2C).

\section{Conclusions and discussion}

We have presented 1.56 $\mu$m spectropolarimetry with high spatial resolution (0.4$''$) and high polarimetric sensitivity (10$^{-4}$ $\langle$I$_\mathrm{c}\rangle$) of the quietest areas of the Sun. A visual inspection of the data revealed that most of the polarisation profiles show clear indications of magnetic substructure within the resolution element, that is, of relative velocities between the Stokes parameters, area, and amplitude asymmetries in Stokes $V$, and/or irregular Stokes $V$ profiles. This substructure can be due to gradients of the physical properties along the LOS (note that Stokes $V$ area asymmetries can only be obtained with vertical gradients) or across the solar surface. Horizontal gradients can be of solar origin (true substructure within the resolution element) or due to stray light in the telescope, or both. To properly take the stray light into account, we need to know the telescope point spread function. At present, this function is not very well characterised, and we preferred to use an approximation of the stray light as an unpolarised quiet-Sun spectrum with a constant contribution in the FOV.

Half of our observed quiet-Sun region is better explained by magnetic substructure within the resolution element. However, we cannot distinguish whether this substructure comes from gradients of the physical parameters along the LOS or from horizontal gradients (across the surface). From a model with two magnetic components (plus a 30\% stray-light contamination), we defined the large (small) population as the one with larger (smaller) filling factor. After a numerical test, we reached the conclusion that the magnetic field strength, inclination, and filling factor of the large component are degenerate and that we can only rely on the magnetic flux density. We also showed that the strength, inclination, and filling factor 
of the small component are constrained by the observations for fields above 250 G: they are isotropic, hG fields. We found a polarity imbalance that occurs for all inclinations. To confirm the solar origin of this imbalance requires more quiet-Sun observations. 

The other half of the observed quiet-Sun area can be explained by a single magnetic field embedded in a non-magnetic atmosphere. From these pixels, 
$\sim$50\% are two-lobed Stokes $V$ profiles. This means that 23\% of the FOV can be adequately explained with a single constant magnetic field embedded in a non-magnetic atmosphere. The others are irregular, very weak profiles, hence the improvement in the fit using a two-component model is negligible. 
From the regular profiles, the magnetic field vector and filling factor were reliable inferred only in 50\%. We inferred hG fields with small filling factors (89\% are below 30\%); at our present spatial resolution, 70\% of the pixel is apparently non-magnetised,
however. There is growing evidence that this large fraction of the resolution element is magnetised, but at much smaller scales than our present resolution capabilities. From the theoretical point of 
view, modern magnetohydrodynamical simulations show a myriad of magnetic fields tangled at scales as small as a few km (e.g. those analysed by Lagg et al. 2016). From the 
observational point of view, each time we improve our resolution capabilities, we detect weaker and smaller-scale fields. In addition, Hanle measurements \citep[e.g.][]{javier_04} claim that there is much more magnetic energy in a pixel than is detected by the Zeeman effect.

\begin{acknowledgements}
The authors are very grateful to an anonymous referee that helped to improve the manuscript and to strengthen the results and 
to L. R. Bellot Rubio for fruitful discussions on the subject of this paper. 
This work is based on observations made with the German GREGOR telescope at the Spanish Observatorio del Teide of the 
Instituto de Astrof\'\i sica de Canarias. The 1.5-meter GREGOR solar telescope was built by the German consortium 
under the leadership of the Kiepenheuer-Institut f\"ur Sonnenphysik with the Leibniz-Institut f\"ur Astrophysik Potsdam, the Institut f\"ur Astrophysik G\"ottingen, the Max-Planck-Institut f\"ur Sonnensystemforschung in G\"ottingen, and the Instituto de Astrof\' isica de Canarias, and with contributions by the Astronomical Institute 
of the Academy of Sciences of the Czech Republic. The GRIS instrument was developed thanks to the support by the Spanish Ministry of Economy and Competitiveness through the project AYA2010-18029 (Solar Magnetism and Astrophysical Spectropolarimetry). Financial support by the Spanish Ministry of Economy and Competitiveness and the 
European FEDER Fund through projects AYA2014-60476-P and AYA2014-60833-P are 
gratefully acknowledged. Financial support by Consolider-Ingenio 2010 CSD2009-00038 is also acknowledged. AAR acknowledges financial support through the Ram\'on y Cajal fellowship. SJGM is grateful for financial support from the Leibniz Graduate School for Quantitative Spectroscopy in Astrophysics, a joint project of the Leibniz Institute for Astrophysics Potsdam and the Institute of Physics and Astronomy of the University of Potsdam.
This paper made use of the IAC Supercomputing facility HTCondor (http://research.cs.wisc.edu/htcondor/).
\end{acknowledgements}

% \bibliographystyle{aa}
% \bibliography{mybib}

\end{document}